\documentclass[aoas,authoryear, preprint]{imsart}
\RequirePackage[OT1]{fontenc}
\RequirePackage{amsthm,amsmath}
\RequirePackage{natbib}
\usepackage{multirow}
\usepackage{amsfonts}
\usepackage{amsmath}
\usepackage{amssymb}
\usepackage{amsthm}
\usepackage{array}
\usepackage{bbm}
\usepackage{bm}
\usepackage{booktabs}
\usepackage[makeroom]{cancel}
\usepackage{color}
\usepackage{float}
\usepackage{enumerate}
\usepackage{graphicx}
\usepackage{mathtools}
\usepackage{newfloat}
\usepackage{tocloft}
\usepackage{wrapfig}

\startlocaldefs
\numberwithin{equation}{section}
\theoremstyle{plain}
\newtheorem{prop}{Proposition}
\endlocaldefs

\begin{document}

\begin{frontmatter}
\title{Dynamic Gene coexpression Analysis with Correlation Modeling}
\runtitle{Dynamic Gene coexpression Analysis}

\begin{aug}
\author{\fnms{Tae} \snm{Kim}\ead[label=e1]{tk382@uchicago.edu}}
\and
\author{\fnms{Dan} \snm{Nicolae}\ead[label=e2]{nicolae@uchicago.edu}}

\affiliation{University of Chicago}
\address{5747 South Ellis Avenue\\
Chicago, IL 60637}
\end{aug}

\begin{abstract}
In many transcriptomic studies, the correlation of genes might fluctuate with quantitative factors such as genetic ancestry. We propose a method that models the covariance between two variables to vary against a continuous covariate. For the bivariate case, the proposed score test statistic is computationally simple and robust to model misspecification of the covariance term. Subsequently, the method is expanded to test relationships between one highly connected gene, such as a transcription factor, and several other genes for a more global investigation of the dynamic of the coexpression network. Simulations show that the proposed method has higher statistical power than alternatives, can be used in more diverse scenarios, and is computationally cheaper. We apply this method to African American subjects from GTEx to analyze the dynamic behavior of their gene coexpression against genetic ancestry and to identify transcription factors whose coexpression with their target genes change with the genetic ancestry. The proposed method can be applied to a wide array of problems that require covariance modeling.
\end{abstract}


\end{frontmatter}
\section{Introduction}
\subsection*{}
Gene coexpression, the covariance structure of gene expression data, measures how genes are functionally connected and provides insights into the design of the transcriptional regulatory system. Ideally, this complicated biological system can be investigated and fully understood through longitudinal observations in multiple and diverse cell types that capture the dynamics of the system. In reality, however, such comprehensive measurements are often unavailable or too expensive, and the expression dynamics must be captured instead through cross-sectional or tissue-specific data sets. In such cases, investigating the dependence structure could be useful. The dependence structure can be especially valuable for characterizing how few key genes are connected to the rest of the transcriptome. For example, we can focus on  transcription factors (TFs) --- genes that help turn transcription of genes on and off --- and study how they are connected to their target genes. To further investigate this problem, we define ``local connectivity" of a transcription factor as its overall connectivity to its target genes.\\

Consider the following biological problem: how does local connectivity vary across various phenotypic conditions?  Past studies have investigated similar problems, such as how subjects in distinct disease groups show distinct coexpression patterns, contributing to a better understanding of disease at a molecular level \citep{de2010differential}. Here, we focus on understanding how coexpression changes with quantitative traits, not discrete conditions such as disease status. In other words, we study the dynamic nature of coexpression. The motivating example of a quantitative trait is genetic ancestry. Ancestry is known to play a critical role in other molecular phenotypes including DNA methylation and gene expression \citep{galanter2017differential, price2008effects}, and so we believe it could play an important role in gene dynamics and gene expression networks as well. In this paper, we study how the local connectivity of candidate genes changes with ancestry. Specifically, we study the gene coexpression of African American subjects to identify candidate transcription factors whose effects on their targets vary with the proportion of African ancestry in their genome. This analysis will lead to a better comprehension of how genes are differentially regulated in distinct populations. \\

The above biological problem can be investigated using multivariate statistical models of gene expression with a covariance structure (characterizing connectivity) that depends on one or more features such as ancestry. This paper focuses on testing the contribution of ancestry on the covariance matrix, and we start from its simplest form by studying the expression levels of two genes. We construct a statistical model that can explain how the correlation varies against genetic ancestry and use that to test if the correlation is constant across conditions. We generalize it to the local connectivity of a transcription factor by meta-analyzing the pairwise statistics. Note that covariance modeling for multivariate data is important in many applications outside the field of genetics. Variance modeling has been widely studied in the context of heteroskedasticity \citep{breusch1979simple, glejser1969new, white1980heteroskedasticity}, and correlation modeling under discrete conditions has been studied in the context of the differential network \citep{ideker2012differential}, but dynamic correlation modeling has been less explored. \\

\cite{li2002genome} and \cite{li2004system} address a similar statistical and scientific problem to ours, using the term ``liquid association" (LA) to conceptualize the evolution of the coexpression pattern for a pair of genes. They analyze the coexpression that changes across different unobserved cellular states that are represented by the expression level of another gene as a proxy. Other studies have built on the liquid association to better identify gene pairs that are likely to be dynamically correlated in genome-scale \citep{yan2017detecting, yu2018new}. However, methods based on liquid association have some limitations. First, they restrict the covariate to be a 1-dimensional vector, and cannot be generalized to more realistic scenarios. Second, they treat the covariate as a random variable that follows a normal distribution, which genetic ancestry does not satisfy. Third, they only test the linear relationship between the covariate and the coexpression. Lastly, the corresponding test statistic does not have a closed-form null distribution and requires a permutation test, leading to computational inefficiency. \\

We apply the classical Rao's score test for heteroskedasticity \citep{breusch1979simple} where the null hypothesis is that the coexpression does not vary with the covariate. This method is generalizable to non-normal, multivariate covariates, and it is also applicable to a non-linear relationship between the variance and the covariate. Moreover, the score test statistic asymptotically follows a chi-squared distribution, and hence it is easily expandable to a large number of tests without excessive computational burden. Subsequently, we tackle the local connectivity problem by expanding the scope of the problem from the relationship of two genes to the relationships between one gene and multiple genes by combining pairwise test statistics. When the number of genes in the local cluster is smaller than the sample size, the desired statistical properties apply to the new combined test statistic as well. \\

The rest of the paper is organized as follows. First, we lay out the framework for the score test that investigates whether the covariance between bivariate normal variables varies against a continuous covariate $X$. Then we propose a way to combine the pair-wise test statistics for one gene and test the global null that the local connectivity of one variable does not change with genetic ancestry. In the simulation section, we show that the proposed method has distinct advantages compared to alternatives such as the likelihood ratio test or liquid association. Finally, we share our real data analysis results using Gene-Tissue Expression (GTEx) data for African American transcriptome and genome. 

\section{Methods}
\subsection{Test for connectivity between two genes\label{sec:framework2}}
Consider the following model (\ref{eq:framework_2genes}) where $\bm{y}_i = \begin{bmatrix} y_{i1} & y_{i2} \end{bmatrix} ^T$ independently follow a bivariate normal distribution for $i = 1, 2, \cdots, N$ with a non-constant covariance that depends on covariate $\bm{x}_i$: 
\begin{equation}
\begin{multlined}
    \begin{bmatrix} y_{i1} \\ y_{i2} \end{bmatrix} = 
    \begin{bmatrix} \bm{z}_i^T \bm{\beta}_1 \\ \bm{z}_i^T \bm{\beta}_2 \end{bmatrix} + 
    \begin{bmatrix} {u_{i1}} \\ u_{i2} \end{bmatrix}\\
    \begin{bmatrix} {u_{i1}} \\ u_{i2} \end{bmatrix} 
    \sim \mathcal{N}\left(
        \begin{bmatrix} 0 \\ 0 \end{bmatrix}, 
        \begin{bmatrix} \sigma_1^2 & \rho(\bm{x}_i) \\ 
        \rho(\bm{x}_i) & \sigma_2^2 \end{bmatrix}
    \right)
    \end{multlined}
    \label{eq:framework_2genes}
\end{equation}
In this model, $\bm{z}_i$ is an $R$-dimensional covariate associated with the mean gene expression level for $N$ individuals and includes an intercept, while $\bm{\beta}_1$ and $\bm{\beta}_2$ are length $R$ vectors for the slope of the mean term. The variances $\sigma_1^2$ and $\sigma_2^2$ are scalars, and $\bm{x}_i$ is a $P$-dimensional covariate that characterizes the effect of $\bm{x}_i$ on the covariance. The function $\rho$ satisfies
\begin{equation}
    \rho(\bm{x}_i) = \rho(\bm{x}_i^T\bm{\alpha} + \alpha_0) \label{eq:rho}
\end{equation}
where $\bm{\alpha}$ is a $P$ dimensional vector and the parameter of interest; $\alpha_0$ is a scalar. The matrices $Z = \{z_{ir}\}_{i=1,r=1}^{N,R}$ and $X = \{x_{ip}\}_{i=1,p=1}^{N,P}$ are both full rank. We aim to test the following null hypothesis about the parameter $\bm{\alpha}$.
\begin{equation}
    H_0: \bm{\alpha} = \bm{0},
    \label{eq:null}
\end{equation}
where all other parameters --- $\alpha_0$, $\bm{\beta}$, $\sigma_1^2$, $\sigma_2^2$ --- are nuisance parameters. Under the null hypothesis, $\rho(\bm{x}_i^T\bm{\alpha} + \alpha_0) = \rho(\alpha_0)$ is constant regardless of $\bm{x}_i$. The model in (\ref{eq:framework_2genes}) offers a flexible framework that can be used for many forms of heteroskedasticity. $\rho$ can take any linear or non-linear form, and only standard assumptions of linearity and additivity in (\ref{eq:rho}) are required. \\


In the context of gene coexpression in admixed populations, $\bm{y}_i$ is gene expression level of an admixed individual $i$ at two genes, and $\bm{x}_i$ is a $P$-dimensional covariate for individual $i$ that holds information about genetic ancestry. It can be a scalar that represents the proportion of ancestry in the genome, a vector of the first few principal components of the genotypes, or a vector of local ancestry at multiple loci. In the application example in section \ref{sec:applications}, global ancestry is used for straightforward interpretability. For this application, $\bm{z}_i$ is equal to $\bm{x}_i$. \\

There are two well-established tools for testing the null hypothesis (\ref{eq:null}): the likelihood ratio test and Rao's score test \citep{breusch1979simple}. Unlike the score test, a likelihood ratio test requires the full specification of the function $\rho$ to estimate the maximum likelihood estimate (MLE) of $\bm{\alpha}$ both under the null hypothesis and under the alternative hypothesis. Possible choices for $\rho$ include sigmoid functions bound to $(-\sqrt{\sigma_1^2\sigma_2^2}$, $\sqrt{\sigma_1^2\sigma_2^2})$ such as logistic function, hyperbolic tangent function, or any cumulative distribution supported on the real line. This modeling strategy leads to two issues: (i) statistical power is sacrificed if $\rho$ is highly mis-specified; (ii) most of the reasonable assumptions for $\rho$, such as the sigmoid functions mentioned above, do not lead to a closed form MLE of $\bm{\alpha}$ under the alternative hypothesis. Numerical optimization of the likelihood leads to computational inefficiency, especially when the test space is large as in our application to gene coexpression. These limitations are also demonstrated in Section \ref{sec:simulations}. \\

On the other hand, Rao's score test requires only the MLE of $\bm{\alpha}$ under the null hypothesis \citep{rao1973linear}. Moreover, under the linear and additive model (\ref{eq:rho}), the test statistic does not depend on the form of $\rho$ while maintaining its asymptotic properties as long as $\rho$ is twice differentiable. In order to test (\ref{eq:null}), we expand the result from \cite{breusch1979simple} to derive a test statistic and its null distribution.

\begin{prop}
In (\ref{eq:framework_2genes}), if the non-diagonal term of $\Sigma$ follows the function $\rho$ as defined in (\ref{eq:rho}), then Rao's score statistic (\ref{eq:q}) does not depend on the unknown function $\rho$ and asymptotically follows $\chi_P^2$ under the null hypothesis (\ref{eq:null}).
\label{prop:q}
\end{prop}

Rao's score statistic is denoted by $q$ and satisfies:
\begin{equation}
\label{eq:q}
    q = \phi^T \Psi^{-1} \phi
\end{equation}
where
$$\phi = 
 \frac{\hat{\rho}(\hat{\sigma}_1^2\hat{\sigma}_2^2-\hat{\rho}^2)\sum \tilde{\bm{x}}_i + (\hat{\sigma}_1^2\hat{\sigma}_2^2+\hat{\rho}^2)\sum_i \hat{u}_{i1}\hat{u}_{i2}\tilde{\bm{x}}_i - \hat{\sigma}_1^2\hat{\rho}\sum \hat{u}_{i2}^2\tilde{\bm{x}}_i - \hat{\sigma}_2^2\hat{\rho}\sum_i \hat{u}_{i1}^2\tilde{\bm{x}}_i}{(\hat{\sigma}_1^2\hat{\sigma}_2^2-\hat{\rho}^2)},$$
$$\Psi =(\hat{\sigma}_1^2\hat{\sigma}_2^2+\hat{\rho}^2)(\hat{\sigma}_1^2\hat{\sigma}_2^2-\hat{\rho}^2)\sum_i \tilde{\bm{x}}_i \tilde{\bm{x}}_i^T -
\frac{4\hat{\sigma}_1^2\hat{\sigma}_2^2\hat{\rho}^2}{N}(\sum \tilde{\bm{x}}_i)(\sum \tilde{\bm{x}}_i^T).$$ 
$\hat{u}_{i1}$ and $\hat{u}_{i2}$ are OLS residuals from (\ref{eq:framework_2genes}), $\hat{\sigma}_1$, $\hat{\sigma}_2$, and $\hat{\rho}$ are maximum likelihood estimates under the null (\ref{eq:null}).
$$\hat{\sigma}_1^2 = \frac{1}{N} \sum_{i=1}^{N} \hat{u}_{i1}^2, \hspace{4mm} \hat{\sigma}_2^2  = \frac{1}{N} \sum_{i=1}^{N} \hat{u}_{i2}^2,\hspace{4mm} \hat{\rho} = \rho(\hat{\alpha}_0) = \frac{1}{N} \sum_{i=1}^{N} \hat{u}_{i1}\hat{u}_{i2}$$
$$\hat{\bm{u}}_{1} = \bm{y}_{1} - Z(Z^TZ)^{-1}Z^T\bm{y}_{1}, \hspace{6mm}\hat{\bm{u}}_{2} = \bm{y}_{2} - Z(Z^TZ)^{-1}Z^T\bm{y}_{2}, \hspace{4mm} \bm{\tilde{x}}_i = \begin{bmatrix} 1 & \bm{x}_i \end{bmatrix}^T$$
The above statistic has two advantages. It does not depend on the unknown function $\rho$, so it is flexible under many shapes of heteroskedasticity. It also has low computational cost; every component of $q$ is easily acquired from the data, and $q$ asymptotically follows $\chi_{P}^2$ \citep{breusch1979simple}.The detailed derivation for Proposition \ref{prop:q} is in Appendix A. \\

The introduced test statistic has convenient asymptotic properties, but the inference is not precise under a finite sample size with the error in the order of $N^{-1}$ \citep{harris1985asymptotic}. Various Monte Carlo experiments have showed that the test rejects the null hypothesis less frequently than indicated by its nominal size \citep{ godfrey1978testing, griffiths1986monte}. In response, corrections have been suggested \citep{cribari2001monotonic, harris1985asymptotic}, and we apply the method from \cite{honda1988size} to ensure the validity of the approximations under smaller sample sizes. Details about the small sample correction are in Appendix B.

\subsection{Test for Local Connectivity \label{sec:frameworkK}} 
Section \ref{sec:framework2} introduced a statistic that tests whether pair-wise correlation changes with the covariate $X$. This section expands the method to more variables to study local connectivity. Consider the extension of (\ref{eq:framework_2genes}) to $K$-dimensional multivariate normal. 
\begin{equation}
\begin{multlined}
    \begin{bmatrix} y_{i1} \\ y_{i2} \\ ... \\ y_{iK} \end{bmatrix} = 
    \begin{bmatrix} \bm{z}_i^T \bm{\beta}_1 \\ \bm{z}_i^T \bm{\beta}_2 \end{bmatrix} + 
    \begin{bmatrix} {u_{i1}} \\ u_{i2} \\ ... \\ u_{iK} \end{bmatrix}\\
    \begin{bmatrix} {u_{i1}} \\ u_{i2} \\ ... \\ u_{iK} \end{bmatrix}
    \sim \mathcal{N}(\bm{0}, \Sigma), \hspace{5mm} 
    \Sigma =  \begin{bmatrix} \sigma_1^2 & \rho_{12}(\bm{x}_i) &...& \rho_{1K}(\bm{x}_i)\\ 
        \rho_{12}(\bm{x}_i) & \sigma_2^2 & ...& \rho_{2K}(\bm{x}_i)\\ 
        ... & ...  & ... & ... \\
        \rho_{K1}(\bm{x}_i) & \rho_{K2}(\bm{x}_i) & ... & \sigma_K^2\end{bmatrix}
    \end{multlined}
    \label{eq:framework_Kgenes}
\end{equation}
$$\rho_{k\ell}(\bm{x}_i) = \rho_{k\ell}(\bm{x}_i^T\bm{\alpha}_{k\ell} + \alpha_{0,k\ell}), \hspace{6mm} k, \ell = 1, \cdots, K, \hspace{5mm} k \neq \ell$$
The global null hypothesis for variable 1 is constructed from the first row of the variance matrix $\Sigma$ of (\ref{eq:framework_Kgenes}).
\begin{equation}
    \bm{H}_0^{(1)}: \bm{\alpha}_{12} = \bm{\alpha}_{13} = \cdots = \bm{\alpha}_{1K} = \bm{0},
\label{eq:globalnull}
\end{equation}
Under $\bm{H}_0^{(1)}$, the correlation of other variable with variable 1 does not change with $X$. We believe testing the global null (\ref{eq:globalnull}) improves statistical power when  ``hot spot"  or ``hub" variables are connected to a lot of other nodes forming cliques or modules. In the context of gene coexpression network, transcription factors (TFs) are considered ``hubs'' that regulate the gene expression of multiple genes. TFs are good candidate genes for the local connectivity analysis because a given TF is likely to affect expression of many of its targets simultaneously. \\

Here, we propose a simple linear combination of pair-wise statistics to test local connectivity of variable 1,
\begin{equation}
d_1 = {q}_{12} + {q}_{13} + \cdots + {q}_{1K} = \sum_{k=2}^{K} {q}_{1k}.
\label{eq:d}
\end{equation}

The statistic $d_1$ in  (\ref{eq:d}) is designed to improve statistical power when pair-wise effect sizes are too small but become substantial when combined. The choice of adding them without additional weighting reflects the lack of prior knowledge about the relationships between the variables. \\

Note that $q_{1k}$ for gene pair $1$ and $k$ can be written as a $\sum_{p=1}^{P} r_{1k,p}^2$ where $r_{1k,p}$ individually follows a standard normal distribution (Appendix C). When expanded to $K$ genes, we can show the following. 
\begin{equation}
\bm{r}_{1,p} = \begin{bmatrix}
r_{12,p} \\ r_{13, p} \\ \cdots \\ r_{1K,p}
\end{bmatrix}  \xrightarrow{d} \mathcal{N}_{K-1}(\bm{0}, H_1) \hspace{5mm} \forall p = 1, \cdots, P
\label{eq:H}
\end{equation}
where $H_1$ is a $(K-1) \times (K-1)$ matrix. Appendix C shows element-wise mapping from $\Sigma$ to $H_1$. Let $H_1 = U_1 \Lambda_1 U_1^T$ be the eigen-decomposition of the covariance matrix $H_1$ in (\ref{eq:H}), where the diagonal matrix $\Lambda$ has eigenvalues $\lambda_{12}, \cdots, \lambda_{1K}$ in a decreasing order. Then, consider the transformation $\bm{r}_{1,p}^* = U\bm{r}_{1,p}$ that follows normal distribution with diagonal covariance matrix $\Lambda_1$. Note that $\|\bm{r}_{1,p}\|_2^2 = \|U\bm{r}_{1,p}\|_2^2 = \|{r_{1k,p}^*}\|_2^2$ due to the orthogonal invariance of $L_2$ norm. Then, $d_1$ asymptotically follows a sum of independent Gamma distributions
\begin{equation}
    d_1 = \sum_{k=2}^{K} q_{1k} \xrightarrow{d} \sum_{k=2}^{K} \Gamma \left( \frac{P}{2}, \frac{\lambda_{1k}}{2}\right)
    \label{eq:d_dist}
\end{equation}

Assuming that we know the true, symmetric, positive definite $H_1$, we can calculate $\lambda_{1k}>0$ for $k = 2, \cdots, K$, and we have expressed the null distribution of $d_1$ as the sum of distributions of gamma variables. We can easily simulate this null distribution. \\ 

\begin{prop}
Under the setting of (\ref{eq:framework_Kgenes}), if none of the variables $y_1, \cdots, y_K$ are perfectly correlated, $d_1$ asymptotically follows the finite sum of Gamma distributions as defined in (\ref{eq:d_dist}) under the global null hypothesis (\ref{eq:globalnull}). 
\label{prop:d}
\end{prop}
A detailed proof of Proposition \ref{prop:d} can be found in Appendix C. \\

The asymptotic distribution of $d_1$ requires knowledge of $\Sigma$, and $\Sigma$ is often unknown in practice, even under the null. When $N$ is sufficiently larger than $K$, the null maximum likelihood estimate $\hat{\Sigma}$ is a consistent estimator for $\Sigma$. Using $\hat{\Sigma}$ guarantees that the test statistic in (\ref{eq:d}) converges in distribution to (\ref{eq:d_dist}). However, when $K$ is larger than $N$, there is no consistent estimator $\Sigma$ that does not require regularization conditions such as sparsity or low-rank. Then, permutation test can be an alternative; testing $Y$ against randomly shuffled $X$ can simulate the null distribution of $d_1$ while preserving the dependence structure. \\

Permutation tests lead to a limited resolution of $p$-values. Imprecise $p$-values prevent accurate inference especially when we need to correct for a large number of hypotheses. Performing a large number of permutations can lead to better resolution of $p$-values but could be computationally wasteful. To strike a balance, we use the sequential precision-improvement permutation test, similar to one suggested by \cite{chen2012exponential}. \\

Precision-improvement permutation test terminates the procedure early if the signal is not strong enough. The detailed procedure is as follows. For every permutation $b = 1, \cdots, B$, we permute the rows (samples) of the covariate matrix $X$ to create $X_b$, so that any existing link between $X$ and $Y$ is broken. Then we compute the degree statistic $d_{kb}$ for gene $k$ using $X_b$ and the data matrix $Y$; $d_{kb}$ should follow the null distribution. Then, $p$-value for $d_k$ is computed as a quantile of $d_k$ compared to the empirical distribution of $d_{kb}$ for each $b$. After the minimum number of permutations pre-defined by the user (100 in the Section \ref{sec:applications}), we count the number of permutations $b$ where $d_{kb}$ is larger than $d_k$. If there are two or more such cases, we terminate the permutation procedure early. Most genes fall into this category leading to $p$-values greater than 0.02. If there are less than 2 such cases observed, we iteratively perform 100 more permutations and re-check the number of $d_{kb}$ with values larger than $d_k$. We repeat until the number of permutations $B$ reaches the predefined maximum number of permutation ($10^6$ in Section \ref{sec:applications}), which is designed to give a good enough resolution of $p$-value given the number of tests that we are performing.

\section{Simulation Studies} \label{sec:simulations}
Here, we evaluate the proposed method through simulations. We focus on the pairwise analysis and compare the performance of the proposed score test with two  alternatives, liquid association and the likelihood ratio test. \\

We sampled $X$ from the univariate standard normal distribution to match the required setting of liquid association. To check the calibration of test statistics under the null hypothesis, we simulated the data matrix $Y$ from 
$$\bm{y}_i \sim \mathcal{N}_2\left(\bm{b}_0 + \bm{z}_i^T\bm{\beta}, \begin{bmatrix} 1 & \bar{\rho} \\ \bar{\rho} & 1 \end{bmatrix}  \right)$$ 
where $\bar{\rho}$ was randomly selected from uniform distribution ranging from -1 and 1, and $\bm{\beta} = \bm{0}$. Different sample sizes were tested to check the behavior of each method under the null hypothesis. For each $N$, we sampled $X$ once, and generate $Y$ 1,000 times. The likelihood ratio test was designed to assume a hyperbolic tangent model for $\rho$,
\begin{align}
    \rho(\bm{\tilde{x}}_i^T\bm{\tilde{\alpha}}) = \frac{e^{\bm{\tilde{x}}_i^T\bm{\tilde{\alpha}}}-1}{e^{\bm{\tilde{x}}_i^T\bm{\tilde{\alpha}}}+1},
    \label{eq:data_generating_fisher}
\end{align}
with $\tilde{\bm{x}}_i = \begin{bmatrix} 1 & \bm{x}_i\end{bmatrix}$, $\tilde{\bm{\alpha}} = \begin{bmatrix} \alpha_0 & \bm{\alpha} \end{bmatrix}$. Note that (\ref{eq:data_generating_fisher}) is the inverse of Fisher transformation, 
$\frac{1}{2} \bm{\tilde{x}}_i^T\bm{\tilde{\alpha}} = \frac{1}{2}log \left(\frac{1+\rho}{1-\rho}\right)$.  We used \textit{optim} function in R to find $\hat{\bm{\alpha}}_{\text{MLE}}$ under the alternative hypothesis. All three methods control the type I error at the nominal size well. The result for $N=70$ is shown in Table \ref{tab:pairwise_simulation}. \\

\begin{figure}
        \includegraphics[width=0.4\textwidth]{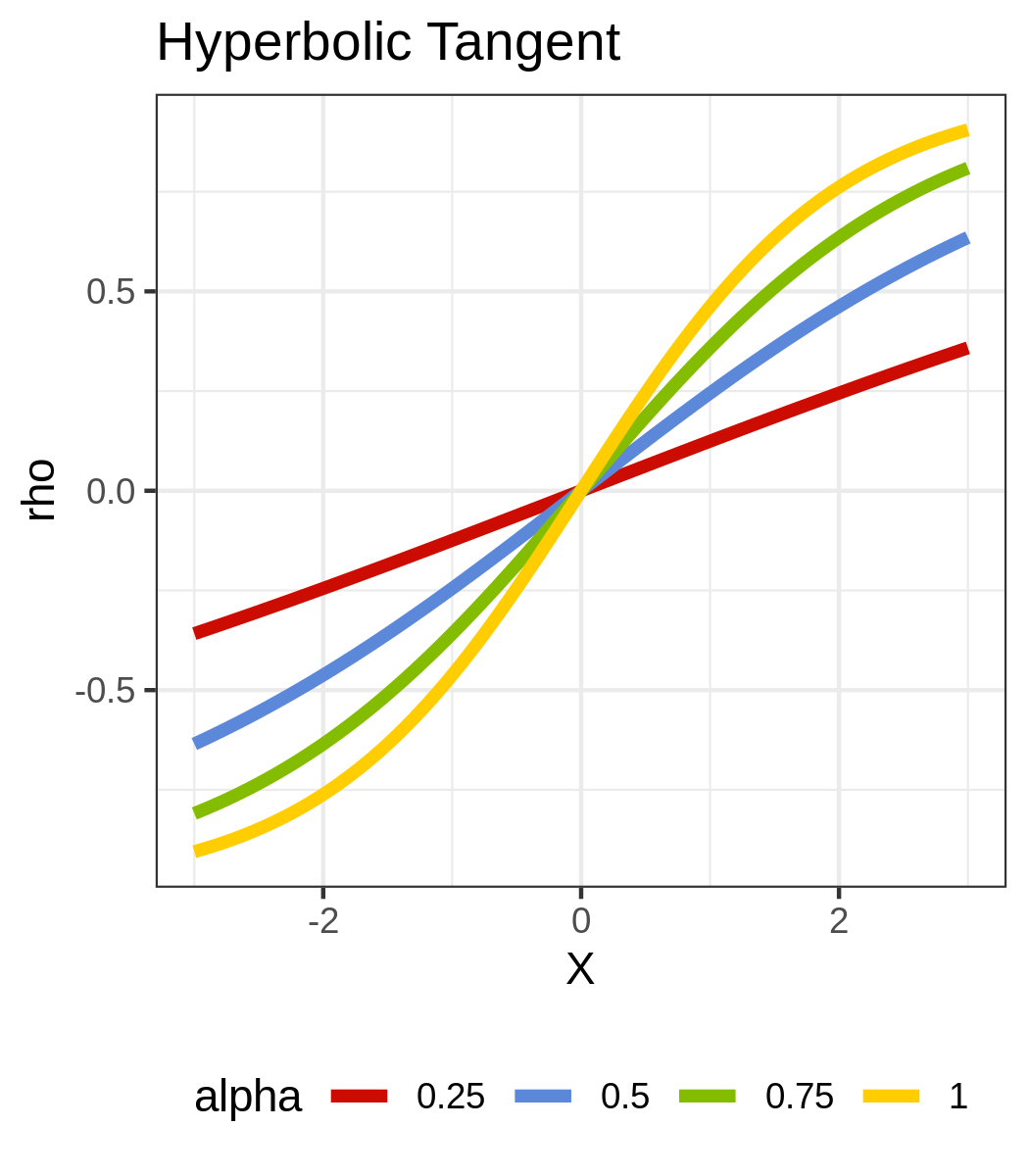}
        \includegraphics[width=0.4\textwidth]{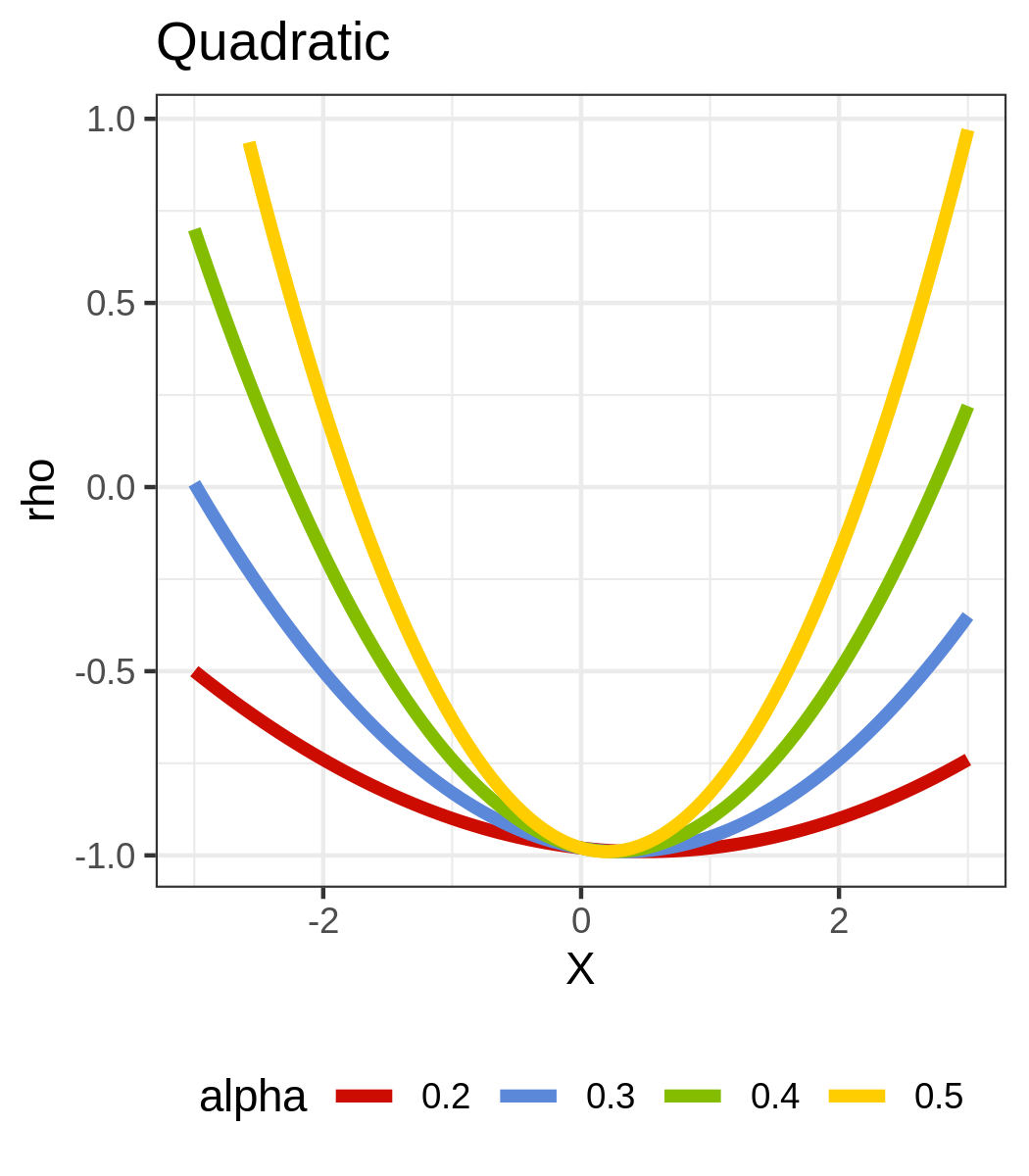}
        \caption{Generation of $\rho$ for the simulations. The two example functions $\rho$ above show the relationship between $X$ and $\rho$ for the two example functions and varying levels of $\alpha$. }
        \label{fig:sim}
\end{figure}

Next, we generated the data under the alternative hypothesis to compare power. We used $N=70$ to reflect the sample size of GTEx data. For $i = 1, \cdots, N$, we generated $\rho(\bm{\tilde{x}}_i^T\bm{\tilde{\alpha}})$ from hyperbolic tangent function in (\ref{eq:data_generating_fisher}). We drew $Y$ from (\ref{eq:framework_2genes}) with varying levels of $\bm{\alpha}$, 1000 times each. The hyperbolic tangent model places the likelihood ratio test at an advantage because the model is correctly specified, so as a contrasting case, we also used a quadratic model to generate $\rho$ as follows,

\begin{align}
    \rho(\alpha_0 + \bm{\tilde{x}}_i^T\bm{\tilde{\alpha}}) = ( - 0.1 + \bm{\tilde{x}}_i^T\bm{\tilde{\alpha}})^2 - 0.99,
    \label{eq:data_generating2_quadratic}
\end{align}

\noindent where subtracting 0.99 is to ensure numerical stability.
Figure \ref{fig:sim} (a) and (b) show the shape of $\rho$ with respect to $X$ with varying levels of $\bm{\alpha}$. \\

Table \ref{tab:pairwise_simulation} summarizes the result.  When $\rho$ is generated from hyperbolic tangent function, likelihood ratio test generally outperforms the other two methods, as expected, since the model is correctly specified in LR test. Score test and liquid association perform similarly. Meanwhile, when $\rho$ is generated from quadratic function, score function clearly outperforms the other two methods. Hence, the simulations show that the proposed score test is robust to the shape of heteroskedasticity. Table \ref{tab:sim}  shows the distribution of computation times of each method to compute the test statistic once in the scale of $log_{10}$ for 1000 simulations under quadratic model with $\alpha = 0.5$. The score test is the most efficient, because the likelihood ratio test requires numerical estimation of MLEs both under the null and under the alternative hypothesis while liquid association requires permutation test for inference. The simulations were done sequentially (in a non-parallel manner) with LAMBDA QUAD workstation with Intel Xeon W-2175 processor.

\begin{table}
\centering 
\begin{tabular}{@{\extracolsep{2pt}} c||c||c||cccc||cccc} 
 & \multicolumn{2}{c}{$\rho$} & \multicolumn{4}{c}{tanh} & \multicolumn{4}{c}{quadratic}\\
\\[-1.8ex]\hline 
\hline \\[-1.8ex]
$\beta$ & $\bm{\alpha}$ & 0 & 0.25 & 0.5 & 0.75 & 1 & 0.2 & 0.3 & 0.4 & 0.5\\ 
\hline \\[-1.8ex]
\multirow{3}{*}{0}& Score & 0.047 & 0.148 & 0.442 & 0.755 & 0.911 &0.912 & 0.831 & 0.735 & 0.699\\
&LA    &0.041 & 0.145 & 0.468 & 0.767 & 0.919 & 0.036 & 0.017 & 0.012 & 0.02\\
&LR    & 0.056 & 0.174 & 0.527 & 0.838 & 0.964 & 0.164 & 0.27 & 0.356 & 0.501\\
\\[-1.8ex]\hline 
\hline \\[-1.8ex]
\multirow{3}{*}{1}& Score & 0.051 & 0.14 & 0.411 & 0.732 & 0.888 & 0.908 & 0.821 & 0.728 & 0.71\\
&LA    & 0.054 & 0.137 & 0.431 & 0.737 & 0.895 & 0.03 & 0.017 & 0.007 & 0.015\\
&LR    & 0.066 & 0.167 & 0.503 & 0.824 & 0.952 & 0.158 & 0.238 & 0.339 & 0.454\\
\\[-1.8ex]\hline 
\hline \\[-1.8ex]
\end{tabular}
 \caption{Proportion of simulations for each method that shows $p$-value $<0.05$ at each data generating model and $\alpha$ level. We use two functions for $\rho$, hyperbolic tangent and quadratic function. The likelihood ratio test was conducted under the assumption that $\rho$ is generated by hyperbolic tangent function. The intercept $\alpha_0$ is 0 in all cases.\label{tab:pairwise_simulation}}
\end{table} 

\begin{table}[!htbp] \centering 
\begin{tabular}{@{\extracolsep{5pt}} ccccccc} 
\\[-1.8ex]\hline 
\hline \\[-1.8ex] 
Test & Min. & 1st Qu. &  Median  &   Mean & 3rd Qu.  &   Max. \\
\hline \\[-1.8ex] 
Score & $2.62 \times 10^{-5}$ & $2.96\times 10^{-5}$ & $3.03\times 10^{-5}$ & $3.13\times 10^{-5}$ & $3.12\times 10^{-5}$ & $2.21\times 10^{-3}$ \\ 
LA & $3.94\times 10^{-3}$ & $4.06\times 10^{-3}$ & $4.22\times 10^{-3}$ & $4.43\times 10^{-3}$ & $4.28\times 10^{-3}$ & $6.96\times 10^{-2}$ \\ 
LR& $3.04\times 10^{-1}$ & $5.32\times 10^{-1}$ & $6.51\times 10^{-1}$ & $6.94\times 10^{-1}$ & $8.05\times 10^{-1}$ & $2.29\times 10^{-1}$\\
\hline\\
\end{tabular} 
  \caption{Distribution of time (in seconds) for getting one score statistic for the each method from 1,000 simulations. The proposed method is faster than the likelihood ratio test in the order of $10^4$, and than the liquid association in the order of $10^2$.  \label{tab:sim} }

\end{table} 

\section{Applications to GTEx Data} \label{sec:applications}
We applied the method to the expression levels in muscle skeletal tissue in GTEx data where 71, highest among all tissues, African American samples are available. We aimed to find transcription factors that change their coexpression pattern with their target genes as the genome's proportion of African ancestry changes. We acquired a list of transcription factors from TF checkpoint database from \cite{chawla2013tfcheckpoint}. We also acquired a list of target genes for each transcription factors from TF2DNA database in \cite{pujato2014prediction}. We only took into consideration target genes with the highest binding scores. We used the proportion of African ancestry for each individual as the covariate, and it is inferred from the genotype data using the software LAMP \citep{pacsaniuc2009imputation}. The proportion of African ancestry levels from the subjects range from 14\% to 96\%. \\

For each of the $k = 1, \cdots, K = 848$ transcription factor encoding gene, we computed the pair-wise statistic $q_{kj}$ for all its target genes $j = 1, \cdots, J_k$, where $J_k$ is the number of target genes for each transcription factor $k$. Then, we computed $d_k = \sum_{j=1}^{J_k} q_{kj}$ to test the hypothesis that the correlation between the transcription factor $k$ and its targets remain the same across different genetic industry. We first divided $d_k$ with the number of targets $J_k$ to compute the average score of all the target genes for the given TF $k$, and we made a heuristic comparison against $\chi_1^2$ distribution. Under the null hypothesis, the expectation of $d_k/J_k$ is 1, although the variance is not trivial due to high dependence. Then, we chose 10 genes with the top $d_{k}/J_k$ values to perform the permutation test. \\

Table \ref{tab:results} summarizes the top 5 transcription factors with the highest average $d_k$ values and their $p$-values computed from sequential permutation tests. The adjusted $p$-values were computed using Benjamini Hochberg procedure. 

\begin{table}[!htbp] \centering 
\begin{tabular}{@{\extracolsep{5pt}} ccc} 
\\[-1.8ex]\hline 
\hline \\[-1.8ex] 
Gene Name & $p$-value & Adjusted $p$-value \\ 
\hline \\[-1.8ex] 
ATOH8 & $1.47 \times 10^{-5}$ & 0.015 \\ 
ZNF678 & $3.26 \times 10^{-5}$ &  0.014\\ 
ZNF638 & $2.15 \times 10^{-4}$ &  0.060\\ 
ZBTB32 & $5.55 \times 10^{-4}$ & 0.118\\
FARSA & $1.53 \times 10^{-3}$ & 0.261\\
\hline \\[-1.8ex] 
\end{tabular} 
\caption{Top 5 transcription factors out of 848 in muscle skeletal tissue. The $p$-values were obtained through the sequential precision-improvement permutation test \label{tab:results}, and adjusted $p$-values were obtained via Benjamini-Hochberg procedure. } 
\end{table} 

\begin{figure}[H]
    \centering
    \includegraphics[width=0.7\textwidth]{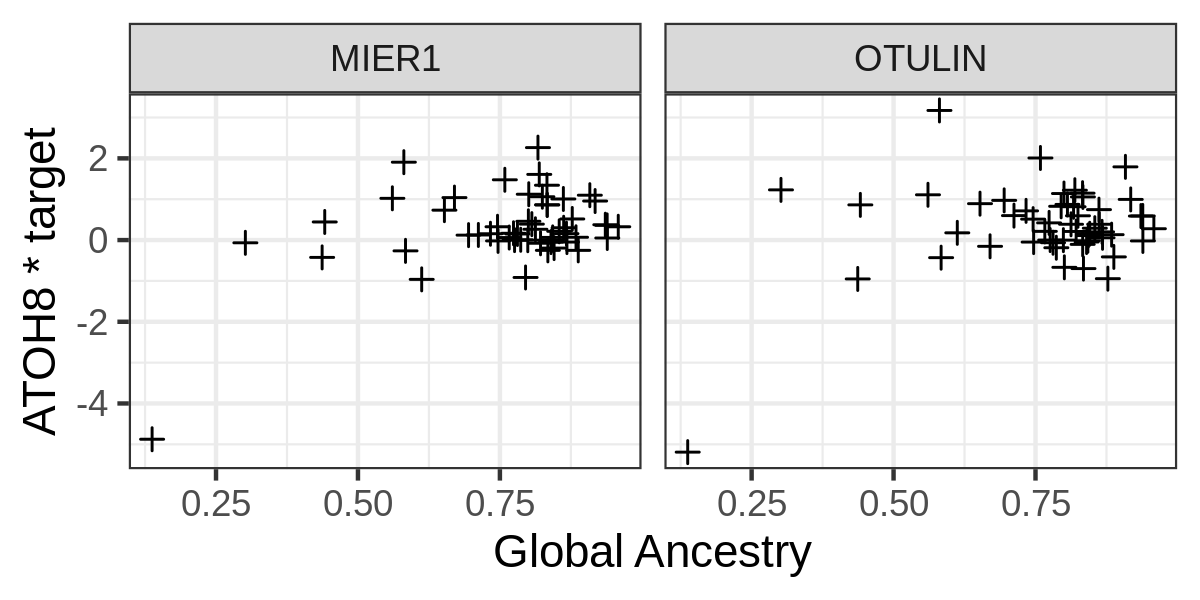}
    \caption{Top two target genes of ATOH8. The product of the expression levels of ATOH8 and two targets with the highest scores (MIER1 and OTULIN) are plotted on the $y$ axis against global ancestry. The product of two genes' expression levels is used as a proxy for the covariance. For both genes, the correlation between two genes become stronger as samples have higher proportion of African ancestry.}
    \label{fig:atoh8}
\end{figure}
\noindent Two genes, ATOH8 and ZNF678, maintain their significance level of 0.05 after adjusting for investigating 848 transcription factors. For the top gene ATOH8, the two highest contributing target genes were MIER1 and OTULIN, each having scores of 58.27 and 29.02 (Figure \ref{fig:atoh8}).\\

\section{Discussion}
We propose a method to test whether the covariance between bivariate normal variables changes with quantitative covariates. We further expanded our scope of analysis by looking at local connectivity --- how one variable's connectivity with multiple other variables change with continuous covariates. We provided a real data example by identifying major transcription factor genes that are differentially connected with their targets by genetic ancestry. \\

Our method is more flexible than  alternatives for covariance testing, but it still has some limitations. First, when there are more variables than the available sample size, as often is the case for many modern data sets, we ultimately turn to a permutation test, not being able to take advantage of the theoretical results. It is challenging, to say the least, to find consistent estimators for the eigenvalues of $\Sigma$ in high-dimensional data, but there could be other ways to estimate $\Sigma$ depending on application. For example, one could impose a sparsity assumption on the covariance matrix, and that can lead to a reliable estimate of $\Sigma$ and subsequently $H$ in (\ref{eq:H}). Such assumption is too restrictive in our context, but in other applications, one can take the liberty to make structural assumptions on the covariance matrix. Second, the score statistic $q$ was derived under the normality assumption of the data set, although simulations show that the result is quite robust to distribution mis-specification. These limitations of the methods propose possible future research topics: (1) better way to estimate $H$ in (\ref{eq:H}) when $K > N$ to preserve asymptotic results, and (2) a non-parametric version of correlation analysis that can generalize to any underlying distributions. \\

In practice, users of this method should note the method's sensitivity to correct mean modeling. For example, if correct data generating procedure is $\bm{z}_i = \bm{x}_i$ + $\bm{x}_i^2$, missing the quadratic term (mis-specifying $\bm{z}_i$ as $\bm{x}_i$) would incorrectly assign the effects of $\bm{x}_i^2$ to the residuals. Then, the score statistic computed from the residuals would show remaining effects from the mean term, and hence the type I error will be inflated. Therefore, the users are expected to be careful in regressing out all the relevant terms so that the mean term effects do not spill over to the variance term.\\ 

Alternative ways have been proposed for some components of our methodology. An important aspect is the formulation of the combined statistic $d$. \cite{chen2012exponential} discusses two ways to construct the alternative hypothesis for testing the global null in (\ref{eq:d}). One way, called a sparse alternative, is to test whether only a small number among all tests have non-zero effects while all other tests are null. Another way is to test if at least one test has a non-zero effect size, as we did in (\ref{eq:d}). Based on prior  knowledge in coexpression network, we assume that there are many small signals instead of few large, so we choose the latter, which is equal to (\ref{eq:d}). Another aspect is about deriving the null distribution of the sum of correlated $\chi_P^2$. \cite{moschopoulos1985distribution} provides another way to derive the distribution of (\ref{eq:d_dist}) by expressing its cumulative distribution in a form of infinite sum, but we believe it is impractical.\\

We believe the proposed method can be applied to data problems in diverse domains, especially where certain central variables are connected to many other variables as in the transcriptional regulatory networks. Many network problems, such as protein interactions, metabolic networks, co-authorship networks, and semantic networks, are known to have, or have something close to, a scale-free topology, indicating that the important variables in those networks can be tested against other variables. The proposed correlation analysis will provide insights into the building blocks of diverse network problems by looking at the pairwise and more than pairwise relationships among the variables.

\pagebreak
 
\section*{Appendix A. Derivation of the test statistic}
Consider the model in (\ref{eq:framework_2genes}). 
\begin{equation*}
    \begin{bmatrix} y_{i1} \\ y_{i2} \end{bmatrix} = 
    \begin{bmatrix} \bm{z}_i^T\bm{\beta}_1 \\ \bm{z}_i^T \bm{\beta}_2 \end{bmatrix} + \begin{bmatrix} u_{i1} \\ u_{i2} \end{bmatrix}
\end{equation*}
\begin{equation*}
    \begin{bmatrix} u_{i1} \\ u_{i2} \end{bmatrix} \sim \mathcal{N} \left( 
    \begin{bmatrix} 0 \\0 \end{bmatrix}, \Sigma_i = \begin{bmatrix} \sigma_1^2 & \rho(\bm{\tilde{x}}_i^T\bm{\tilde{\alpha}}) \\ \rho(\bm{\tilde{x}}_i^T\bm{\tilde{\alpha}}) & \sigma_2^2 \end{bmatrix}\right)
\end{equation*}
where $\tilde{\bm{\alpha}} = \begin{bmatrix} \alpha_0& \bm{\alpha}^T \end{bmatrix}$, and $\tilde{\bm{x}_i}$ is $\begin{bmatrix} 1 & \bm{x}_i\end{bmatrix}^T$. Let $\ell(\theta)$ be the log likelihood depending on a vector of parameters $\theta^T = \begin{bmatrix}\bm{\beta}_1 &  \bm{\beta}_2 &\sigma_1^2&\sigma_2^2  & \alpha_0&\bm{\alpha}\end{bmatrix}$, with $d = \frac{\partial \ell}{\partial \theta}$ as the first derivative (score) vector and $\mathcal{I} = -E(\partial^2 \ell / \partial \theta \partial \theta^T)$ as the information matrix. Following \cite{rao1973linear}, the score statistic for testing the null hypothesis $\bm{\alpha} = 0$ is 
$$q = \hat{d}^T \hat{\mathcal{I}}^{-1} \hat{d}$$
where the hats indicate that the score $d$ and Fisher Information $I$ are evaluated with $\hat{\theta}$, the restricted maximum likelihood estimate: 
$$\hat{\theta} =  \begin{bmatrix}
\hat{\bm{\beta}}_1 & \hat{\bm{\beta}}_2 & \hat{\sigma}_1^2 & \hat{\sigma}_2^2 & \hat{\alpha}_0 & \bm{0} 
\end{bmatrix},
$$
where $\hat{\bm{\beta}}_1$ and $\hat{\bm{\beta}}_2$ are linear regression coefficients. The error term $u$ can be replaced with the OLS residuals $\hat{u}$. Then, $\hat{\sigma}_1^2 = \sum_{i=1}^{N} u_{i1}^2/N$ and $\hat{\sigma}_2^2 = \sum_{i=1}^{N} u_{i2}^2/N$. $\hat{\alpha}_0$ is a value that makes $\rho(\hat{\alpha}_0) = \hat{\rho} = \sum_{i=1}^{N} u_{i1}u_{i2}/N$.\\

Consider partitioning $\theta$ into $\begin{bmatrix} \theta_1; & \theta_2\end{bmatrix} = \begin{bmatrix} \bm{\beta}_1 & \bm{\beta}_2; & \sigma_1^2 & \sigma_2^2 & \alpha_0&\bm{\alpha}\end{bmatrix}$. Then, the constraints refer to only the second subset $\theta_2$. The score vector $d$ can be partitioned similarly as $d = \begin{bmatrix} d_1 & d_2 \end{bmatrix}$. Furthermore, the information matrix is block diagonal between $\theta_1$ and $\theta_2$, so that $\mathcal{I}_{21} = -E(\partial^2\ell / \partial \theta_2 \partial \theta_1^T) = 0$. $\hat{d}_1$ is $\bm{0}$. Moreover, $\hat{d}_1^T\hat{\mathcal{I}}_{11}^{-1} \hat{d}_1$ is evaluated at the MLEs of $\theta_1$ and hence is 0. Therefore, the score statistic is
$$q = \hat{d}_2^T \hat{\mathcal{I}}_{22}^{-1} \hat{d}_2.$$

\subsection*{Score vector}
This section derives the score vector with respect to $\sigma_1^2$, $\sigma_2^2$, $\tilde{\bm{\alpha}}$. For the variance parameter, for example $\tilde{\bm{\alpha}}$, the derivative of the log likelihood is as follows:
$$\frac{\partial \ell} {\partial \tilde{\bm{\alpha}}} =
-\frac{1}{2} \sum_{i=1}^{N} \left(
\frac{\partial log|\Sigma_i|}{\partial\tilde{\bm{\alpha}}} + 
\frac{\partial \bm{u}_i^T \Sigma^{-1}\bm{u}_i}{\partial \tilde{\bm{\alpha}}}
\right)
$$
First term:
\begin{align*}
    \frac{\partial log(|\Sigma_i|)}{\partial \sigma_1^2}
 &= \frac{\sigma_2^2}{\sigma_1^2\sigma_2^2 - \rho_{i}^2}\\
     \frac{\partial log(|\Sigma_i|)}{\partial \sigma_2^2}
 &= \frac{\sigma_1^2}{\sigma_1^2\sigma_2^2 - \rho_{i}^2}\\
     \frac{\partial log(|\Sigma_i|)}{\partial \tilde{\bm{\alpha}}}
 &= \frac{-2\rho_{i}\rho_{i}'}{\sigma_1^2\sigma_2^2 - \rho_{i}^2}\bm{\tilde{x}}_i
\end{align*}
where $\frac{\partial \rho(\tilde{\bm{\alpha}}^T\tilde{\bm{x}}_i)}{\partial \tilde{\bm{\alpha}}} = \rho_i' \tilde{\bm{x}}_i$ with some scalar $\rho_i'$.

\begin{align*}
\bm{u}_i^T \Sigma_i^{-1}\bm{u}_i&=\frac{1}{\sigma_1^2 \sigma_2^2 - \rho_{i}^2}\begin{bmatrix}u_{i1} & u_{i2} \end{bmatrix}  \begin{bmatrix} \sigma_2^2 & -\rho_{i} \\ -\rho_{i} & \sigma_1^2 \end{bmatrix}  \begin{bmatrix}u_{i1}\\ u_{i2} \end{bmatrix}\\
&= \frac{1}{\sigma_1^2 \sigma_2^2 - \rho_{i}^2} \begin{bmatrix} u_{i1} \sigma_2^2 - u_{i2} \rho_{i} & -u_{i1}\rho_{i} + u_{i2}\sigma_1^2 \end{bmatrix} \begin{bmatrix} u_{i1} \\ u_{i2} \end{bmatrix} \\
	&= \frac{1}{\sigma_1^2\sigma_2^2 - \rho_{i}^2} \left(
	u_{i1}^2\sigma_2^2 - u_{i1}u_{i2}\rho_{i} - u_{i1}u_{i2}\rho_{i} + u_{i2}^2\sigma_1^2
	\right)\\
	&= \frac{1}{\sigma_1^2\sigma_2^2 - \rho_{i}^2} ( u_{i1}^2\sigma_2^2 + u_{i2}^2\sigma_1^2 - 2u_{i1}u_{i2}\rho_{i})
\end{align*}

\begin{align*} \frac{\partial \bm{u}_i^T \Sigma^{-1}\bm{u}_i}{\partial \sigma_1^2}
&= 
\frac{
	u_{i2}^2(\sigma_1^2\sigma_2^2 - \rho_{i}^2) - (u_{i2}^2\sigma_1^2 + u_{i1}^2\sigma_2^2 - 2u_{i1}u_{i2}\rho_{i})\sigma_2^2
}{
	(\sigma_1^2\sigma_2^2 - \rho_{i}^2)^2
}\\
  \frac{\partial \bm{u}_i^T \Sigma^{-1}\bm{u}_i}{\partial \sigma_2^2}
  &= 
 \frac{
 u_{i1}^2(\sigma_1^2\sigma_2^2 - \rho_{i}^2) - (u_{i2}^2\sigma_1^2 + u_{i1}^2\sigma_2^2 - 2u_{i1}u_{i2}\rho_{i})\sigma_1^2
 }{
 (\sigma_1^2\sigma_2^2 - \rho_{i}^2)^2
 }\\
  \frac{\partial \bm{u}_i^T \Sigma^{-1}\bm{u}_i}{\partial \tilde{\bm{\alpha}}}
  &= 
\frac{-2u_{i1}u_{i2}\rho_{i}'(\sigma_1^2\sigma_2^2 - \rho_{i}^2) - (u_{i2}^2\sigma_1^2 + u_{i1}^2\sigma_2^2 - 2u_{i1}u_{i2}\rho_{i})(-2\rho_{i})(\rho_{i}')
 }{
 (\sigma_1^2\sigma_2^2 - \rho_{i}^2)^2
 }\tilde{\bm{x}}_i\\
 &= \frac{u_{i1}u_{i2}(-2\rho_{i}'(\sigma_1^2\sigma_2^2 - \rho_{i}^2) - 4\rho_{i}^2\rho_{i}') + (2u_{i2}^2\sigma_1^2\rho_{i} + 2u_{i1}^2\sigma_2^2\rho_{i})\rho_{i}'}{(\sigma_1^2\sigma_2^2 - \rho_{i}^2)^2} \tilde{\bm{x}}_i \\
 &=2\rho_{i}' \frac{u_{i2}^2\sigma_1^2\rho_{i} + u_{i1}^2\sigma_2^2\rho_{i} - u_{i1}u_{i2}(\sigma_1^2\sigma_2^2 - \rho_{i}^2 + 2\rho_{i}^2)}{(\sigma_1^2\sigma_2^2 - \rho_{i}^2)^2}\tilde{\bm{x}}_i
 \end{align*}
 
 Plugging in the MLEs for each nuisance parameter, we can replace $$\sigma_1^2 \text{  with  }\hat{\sigma}_2^2 = \sum_{i=1}^{N} \hat{u}_{i1}^2/N$$
 $$\sigma_2^2 \text{    with    } \hat{\sigma}_2^2 = \sum_{i=1}^{N} \hat{u}_{i2}^2/N$$
 $$\tilde{\bm{\alpha}}\text{    with    } \hat{\tilde{\bm{\alpha}}} = (\hat{\alpha}_0, \bm{0})$$
$$\rho \text{    with    }  \hat{\rho} = \sum_{i=1}^{N} \hat{u}_{i1}\hat{u}_{i2}/N$$ 
$$\rho' \text{    with    } \hat{\rho}' = \frac{\partial \rho(\tilde{\bm{\alpha}}^T\tilde{\bm{x}}_i)}{\partial \tilde{\bm{\alpha}}} \mid_{\theta = \hat{\theta}}$$

 \begin{align*}
 \hat{d}_{\sigma_1^2} = \frac{\partial \ell}{\partial \sigma_1^2}\mid_{\theta = \hat{\theta}}&= 
 -\frac{1}{2} \sum_{i=1}^{N} \left(
 \frac{\hat{\sigma}_2^2}{\hat{\sigma}_1^2\hat{\sigma}_2^2-\hat{\rho}^2} + \frac{\hat{u}_{i2}^2(\hat{\sigma}_1^2\hat{\sigma}_2^2-\hat{\rho}^2) - (\hat{u}_{i2}^2\hat{\sigma}_1^2 + \hat{u}_{i1}^2\hat{\sigma}_2^2-2\hat{u}_{i1}\hat{u}_{i2}\hat{\rho})\hat{\sigma}_2^2}{(\hat{\sigma}_1^2\hat{\sigma}_2^2-\hat{\rho}^2)^2}
 \right)
\\
&= -\frac{1}{2}\left( \frac{\hat{\sigma}_2^2N}{\hat{\sigma}_1^2\hat{\sigma}_2^2-\hat{\rho}^2} + \frac{\sum \hat{u}_{i2}^2(\hat{\sigma}_1^2\hat{\sigma}_2^2-\hat{\rho}^2) - (\hat{\sigma}_1^2\sum_i \hat{u}_{i2}^2 + \hat{\sigma}_2^2\sum \hat{u}_{i1}^2 - 2\hat{\rho}\sum \hat{u}_{i1}\hat{u}_{i2})\hat{\sigma}_2^2}{(\hat{\sigma}_1^2\hat{\sigma}_2^2-\hat{\rho}^2)^2}\right)\\
&= -\frac{N}{2} \left(
\frac{\hat{\sigma}_2^2(\hat{\sigma}_1^2\hat{\sigma}_2^2-\hat{\rho}^2)}{(\hat{\sigma}_1^2\hat{\sigma}_2^2-\hat{\rho}^2)^2} + \frac{\hat{\sigma}_2^2(\hat{\sigma}_1^2\hat{\sigma}_2^2-\hat{\rho}^2) - (\hat{\sigma}_1^2\hat{\sigma}_2^2+\hat{\sigma}_1^2\hat{\sigma}_2^2-2\hat{\rho}^2))}{(\hat{\sigma}_1^2\hat{\sigma}_2^2-\hat{\rho}^2)^2}
\right)\\
&= -\frac{\hat{\sigma}_2^2N}{2}\left( \frac{(\hat{\sigma}_1^2\hat{\sigma}_2^2-\hat{\rho}^2) + (\hat{\sigma}_1^2\hat{\sigma}_2^2-\hat{\rho}^2) - (2\hat{\sigma}_1^2\hat{\sigma}_2^2-2\hat{\rho}^2)}{(\hat{\sigma}_1^2\hat{\sigma}_2^2-\hat{\rho}^2)^2}\right)\\
&= 0
 \end{align*}
Similarly, $\hat{d}_{\sigma_2^2} = 0$.
    
\begin{align*}
\hat{d}_{\tilde{\bm{\alpha}}} &= -\frac{1}{2} \sum_{i=1}^{N} \left(
\frac{-2\hat{\rho}\hat{\rho}'}{\hat{\sigma}_1^2\hat{\sigma}_2^2-\hat{\rho}^2} + \frac{2\hat{u}_{i2}^2\hat{\sigma}_1^2\hat{\rho}+2\hat{u}_{i1}^2\hat{\sigma}_2^2\hat{\rho}-2\hat{u}_{i1}\hat{u}_{i2}(\hat{\sigma}_1^2\hat{\sigma}_2^2-3\hat{\rho}^2)}{(\hat{\sigma}_1^2\hat{\sigma}_2^2-\hat{\rho}^2)^2}\hat{\rho}'
\right)\tilde{\bm{x}}_i \\
&= \frac{\hat{\rho}\hat{\rho}'\sum\tilde{\bm{x}}_i}{\hat{\sigma}_1^2\hat{\sigma}_2^2-\hat{\rho}^2} - \frac{\hat{\sigma}_1^2\hat{\rho}\hat{\rho}'\sum \hat{u}_{i2}^2\tilde{\bm{x}}_i + \hat{\sigma}_2^2\hat{\rho}\hat{\rho}'\sum_i \hat{u}_{i1}^2\tilde{\bm{x}}_i - (\hat{\sigma}_1^2\hat{\sigma}_2^2\hat{\rho}'+\hat{\rho}^2\hat{\rho}')\sum \hat{u}_{i1}\hat{u}_{i2}\tilde{\bm{x}}_i}{(\hat{\sigma}_1^2\hat{\sigma}_2^2-\hat{\rho}^2)^2}\\
 &= \hat{\rho}' \cdot \frac{(\hat{\sigma}_1^2\hat{\sigma}_2^2\hat{\rho}-\hat{\rho}^3) \sum \tilde{\bm{x}}_i - \hat{\sigma}_1^2\hat{\rho}\sum \hat{u}_{i2}^2\tilde{\bm{x}}_i - \hat{\sigma}_2^2\hat{\rho}\sum_i \hat{u}_{i1}^2\tilde{\bm{x}}_i + (\hat{\sigma}_1^2\hat{\sigma}_2^2+\hat{\rho}^2)\sum_i \hat{u}_{i1}\hat{u}_{i2}\tilde{\bm{x}}_i}{(\hat{\sigma}_1^2\hat{\sigma}_2^2-\hat{\rho}^2)^2}\\
 &= \hat{\rho}' \cdot \left(
 \frac{\hat{\rho}(\hat{\sigma}_1^2\hat{\sigma}_2^2-\hat{\rho}^2)\sum \tilde{\bm{x}}_i + (\hat{\sigma}_1^2\hat{\sigma}_2^2+\hat{\rho}^2)\sum_i \hat{u}_{i1}\hat{u}_{i2}\tilde{\bm{x}}_i - \hat{\sigma}_1^2\hat{\rho}\sum \hat{u}_{i2}^2\tilde{\bm{x}}_i - \hat{\sigma}_2^2\hat{\rho}\sum_i \hat{u}_{i1}^2\tilde{\bm{x}}_i}{(\hat{\sigma}_1^2\hat{\sigma}_2^2-\hat{\rho}^2)^2}
 \right)
 \end{align*}  
Consider another partitioning into 
$\hat{d}_{\tilde{\bm{\alpha}}} = \begin{bmatrix}
\hat{d}_{\alpha_0}; & \hat{d}_{\bm{\alpha}}
\end{bmatrix}$. Below, we show that $\hat{d}_{\alpha_0} = 0$.
\begin{align*}
&\hat{\rho}(\hat{\sigma}_1^2\hat{\sigma}_2^2 - \hat{\rho}^2)\sum_{i=1}^{N} 1 + (\hat{\sigma}_1^2\hat{\sigma}_2^2+\hat{\rho}^2)\sum_i \hat{u}_{i1}\hat{u}_{i2} - \hat{\sigma}_1^2\hat{\rho}\sum \hat{u}_{i2}^2 - \hat{\sigma}_2^2\hat{\rho} \sum_i \hat{u}_{i1}^2\\
&= \hat{\rho}(\hat{\sigma}_1^2\hat{\sigma}_2^2-\hat{\rho}^2)N + (\hat{\sigma}_1^2\hat{\sigma}_2^2+\hat{\rho}^2)(N\hat{\rho}) - \hat{\sigma}_1^2\hat{\rho}(N\hat{\sigma}_2^2) - \hat{\sigma}_2^2\hat{\rho} (N\hat{\sigma}_1^2)\\
&= 2N\hat{\rho}(\hat{\sigma}_1^2\hat{\sigma}_2^2) - 2N\hat{\rho}\hat{\sigma}_1^2\hat{\sigma}_2^2\\
&= 0
\end{align*}

\subsection*{Fisher Information}
The Fisher information for two variance parameters, for example $\sigma_1^2$ and $\sigma_2^2$, is
$$\mathcal{I}_{\sigma_1^2, \sigma_2^2} = \frac{1}{2} tr \left( \Sigma^{-1} \frac{\partial \Sigma}{\partial \sigma_1^2} \Sigma^{-1} \frac{\partial \Sigma}{\partial \sigma_2^2}\right)$$
\begin{align*}
    \mathcal{I}(\theta)_{\sigma_1^2\sigma_1^2} &= \frac{1}{2}\sum_i  tr \left( \frac{1}{(\sigma_1^2 \sigma_2^2 - \rho_{i}^2)^2} \begin{bmatrix} \sigma_2^2 & -\rho_{i} \\ -\rho_{i} & \sigma_1^2 \end{bmatrix} 
    \begin{bmatrix} 1 & 0 \\ 0 & 0 \end{bmatrix}\begin{bmatrix} \sigma_2^2 & -\rho_{i} \\ -\rho_{i} & \sigma_1^2 \end{bmatrix}
    \begin{bmatrix} 1 & 0 \\ 0 & 0 \end{bmatrix}\right)\\
    &= \frac{1}{2} tr \left( \frac{1}{(\sigma_1^2 \sigma_2^2 - \rho_{i}^2)^2} \begin{bmatrix} \sigma_2^2 & 0 \\ -\rho_{i} & 0 \end{bmatrix}^2 \right)\\
    &= \frac{1}{2} \frac{\sigma_2^4}{(\sigma_1^2\sigma_2^2 - \rho_{i}^2)^2}
\end{align*}
\begin{align*}
    \mathcal{I}(\theta)_{\sigma_2^2\sigma_2^2} &= \frac{1}{2} \sum_i tr \left( \frac{1}{(\sigma_1^2 \sigma_2^2 - \rho_{i}^2)^2} \begin{bmatrix} \sigma_2^2 & -\rho_{i} \\ -\rho_{i} & \sigma_1^2 \end{bmatrix} 
    \begin{bmatrix} 0 & 0 \\ 0 & 1 \end{bmatrix}\begin{bmatrix} \sigma_2^2 & -\rho_{i} \\ -\rho_{i} & \sigma_1^2 \end{bmatrix}
    \begin{bmatrix} 0 & 0 \\ 0 & 1 \end{bmatrix}\right)\\
    &= \frac{1}{2} tr \left( \frac{1}{(\sigma_1^2 \sigma_2^2 - \rho_{i}^2)^2} \begin{bmatrix} 
    0 & -\rho_{i} \\ 0 & \sigma_1^2
    \end{bmatrix}^2 \right)\\
    &= \frac{1}{2} \frac{\sigma_1^4}{(\sigma_1^2\sigma_2^2 - \rho_{i}^2)^2}
\end{align*}

\begin{align*}
\mathcal{I}(\theta)_{\sigma_1^2\sigma_2^2} &= \frac{1}{2} \sum_i tr \left( \frac{1}{(\sigma_1^2 \sigma_2^2 - \rho_{i}^2)^2} \begin{bmatrix} \sigma_2^2 & -\rho_{i} \\ -\rho_{i} & \sigma_1^2 \end{bmatrix} 
\begin{bmatrix} 1 & 0 \\ 0 & 0 \end{bmatrix}\begin{bmatrix} \sigma_2^2 & -\rho_{i} \\ -\rho_{i} & \sigma_1^2 \end{bmatrix}
\begin{bmatrix} 0 & 0 \\ 0 & 1 \end{bmatrix}\right)\\
&= \frac{1}{2} tr \left( \frac{1}{(\sigma_1^2 \sigma_2^2 - \rho_{i}^2)^2} \begin{bmatrix} 
\sigma_2^2 &  0\\ -\rho_{i} & 0
\end{bmatrix} 
\begin{bmatrix} 
0 & \rho_{i}  \\ 0 & -\sigma_1^2
\end{bmatrix} 
\right)\\
&= \frac{1}{2} \sum_i \left(
\frac{1}{(\sigma_1^2\sigma_2^2 - \rho_{i}^2)^2} \begin{bmatrix}
0 & \sigma_2^2\rho_{i} \\ 0 & -\rho_{i}^2
\end{bmatrix}
\right)\\
&= \frac{1}{2} \sum_i \frac{-\rho_{i}^2}{(\sigma_1^2\sigma_2^2 - \rho_{i}^2)^2}
\end{align*}

\begin{align*}
    \mathcal{I}(\theta)_{\tilde{\bm{\alpha}}\tilde{\bm{\alpha}}} &= \frac{1}{2}\sum_i  tr \left( \frac{1}{(\sigma_1^2 \sigma_2^2 - \rho_{i}^2)^2} \begin{bmatrix} \sigma_2^2 & -\rho_{i} \\ -\rho_{i} & \sigma_1^2 \end{bmatrix} 
    \begin{bmatrix} 0 & \rho_{i}' \\ \rho_{i}' & 0 \end{bmatrix}\begin{bmatrix} \sigma_2^2 & -\rho_{i} \\ -\rho_{i} & \sigma_1^2 \end{bmatrix} 
    \begin{bmatrix} 0 & \rho_{i}' \\ \rho_{i}' & 0 \end{bmatrix}  \right)\tilde{\bm{x}}_i \tilde{\bm{x}}_i^T\\
    &= \frac{1}{2} \sum_i tr \left( \frac{1}{(\sigma_1^2 \sigma_2^2 - \rho_{i}^2)^2}
    \begin{bmatrix}
    -\rho_{i}\rho_{i}' & \rho_{i}'\sigma_2^2 \\
    \sigma_1^2\rho_{i}' & -\rho_{i}\rho_{i}'
    \end{bmatrix}^2\right)\tilde{\bm{x}}_i \tilde{\bm{x}}_i^T\\
    &= \frac{1}{2} \sum_{i} tr \left(
    \frac{1}{(\sigma_1^2\sigma_2^2 - \rho_{i}^2)^2} \begin{bmatrix}
    \rho_{i}^2\rho_{i}'^2 & -2\sigma_2^2\rho_{i}^2\rho_{i}'\\
    -2\sigma_1^2\rho_{i}^2\rho_{i}' & \rho_{i}^2\rho_{i}'^2
    \end{bmatrix}
    \right)
    \tilde{\bm{x}}_i\tilde{\bm{x}}_i^T
    \\
    &= \sum_{i=1}^{N}\frac{\rho_{i}^2 \rho_{i}'^2 + \sigma_1^2\sigma_2^2\rho_{i}'^2}{(\sigma_1^2\sigma_2^2 - \rho_{i}^2)^2} \tilde{\bm{x}}_i \tilde{\bm{x}}_i^T 
\end{align*}

\begin{align*}
    \mathcal{I}(\theta)_{\sigma_1^2\tilde{\bm{\alpha}}} &= \frac{1}{2}\sum_i  tr \left( \frac{1}{(\sigma_1^2 \sigma_2^2 - \rho_{i}^2)^2} \begin{bmatrix} \sigma_2^2 & -\rho_{i} \\ -\rho_{i} & \sigma_1^2 \end{bmatrix} 
    \begin{bmatrix} 1 & 0 \\ 0 & 0 \end{bmatrix}\begin{bmatrix} \sigma_2^2 & -\rho_{i} \\ -\rho_{i} & \sigma_1^2 \end{bmatrix}
    \begin{bmatrix} 0 & \rho_{i}' \\ \rho_{i}' & 0 \end{bmatrix}\right)\tilde{\bm{x}}_i\\
    &= -\sum_i \frac{\rho_{i} \rho_{i}'\sigma_2^2}{(\sigma_1^2\sigma_2^2 - \rho_{i}^2)^2} \tilde{\bm{x}}_i
\end{align*}

\begin{align*}
    \mathcal{I}(\theta)_{\sigma_2^2\tilde{\bm{\alpha}}} &= \frac{1}{2} \sum_i tr \left( \frac{1}{(\sigma_1^2 \sigma_2^2 - \rho_{i}^2)^2} \begin{bmatrix} \sigma_2^2 & -\rho_{i} \\ -\rho_{i} & \sigma_1^2 \end{bmatrix} 
    \begin{bmatrix} 0 & 0 \\ 0 & 1 \end{bmatrix}\begin{bmatrix} \sigma_2^2 & -\rho_{i} \\ -\rho_{i} & \sigma_1^2 \end{bmatrix}
    \begin{bmatrix} 0 & \rho_{i}' \\ \rho_{i}' & 0 \end{bmatrix}\right)\tilde{\bm{x}}_i\\
    &= -\sum_i \frac{\rho_{i} \rho_{i}'\sigma_1^2}{(\sigma_1^2\sigma_2^2 - \rho_{i}^2)^2} \tilde{\bm{x}}_i
\end{align*}
Putting them altogether, $\mathcal{I}(\sigma_1^2, \sigma_2^2, \tilde{\bm{\alpha}}) = $
$$\sum_{i=1}^{N} \frac{1}{(\sigma_1^2\sigma_2^2 - \rho_{i}^2)^2}\begin{bmatrix}
\frac{1}{2} \sigma_2^4 & -\frac{1}{2} \rho_{i}^2 & -\rho_{i}\rho_{i}'\sigma_2^2 \tilde{\bm{x}}_i^T\\
-\frac{1}{2} \rho_{i}^2 & \frac{1}{2} \sigma_1^4 & -\rho_{i} \rho_{i}'\sigma_1^2 \tilde{\bm{x}}_i^T \\
-\rho_{i} \rho_{i}'\sigma_2^2 \tilde{\bm{x}}_i & -\rho_{i} \rho_{i}'\sigma_1^2 \tilde{\bm{x}}_i &
(\rho_{i}^2 \rho_{i}'^2 + \sigma_1^2\sigma_2^2 \rho_{i}'^2)\tilde{\bm{x}}_i \tilde{\bm{x}}_i^T
\end{bmatrix}
$$
Replacing the nuisance parameters with respective MLES ($\sigma_1^2$ to $\hat{\sigma}_1^2$, and so on; $\tilde{\bm{\alpha}} = (\hat{\alpha}_0, \alpha)$):  the Fisher information matrix is
$$\tilde{\mathcal{I}} = \frac{1}{(\hat{\sigma}_1^2\hat{\sigma}_2^2-\hat{\rho}^2 )^2}
\begin{bmatrix}
\frac{\hat{\sigma}_2^4N}{2} & -\frac{\hat{\rho}^2N}{2} & -\hat{\sigma}_2^2\hat{\rho}\hat{\rho}'\sum_i \tilde{\bm{x}}_i \\
-\frac{\hat{\rho}^2N}{2} & \frac{\hat{\sigma}_1^4N}{2} & -\hat{\sigma}_1^2\hat{\rho}\hat{\rho}' \sum_i \tilde{\bm{x}}_i\\
-\hat{\sigma}_2^2\hat{\rho}\hat{\rho}'\sum_i \tilde{\bm{x}}_i & -\hat{\sigma}_1^2\hat{\rho}\hat{\rho}' \sum_i \tilde{\bm{x}}_i & (\hat{\rho}^2\hat{\rho}'^2+\hat{\sigma}_1^2\hat{\sigma}_2^2\hat{\rho}'^2)\sum_i \tilde{\bm{x}}_i\tilde{\bm{x}}_i^T
\end{bmatrix}$$
Consider the above matrix as a 2 by 2 block matrix of $\begin{bmatrix} A & B \\ C & D \end{bmatrix}$, where $A = \frac{1}{(\hat{\sigma}_1^2\hat{\sigma}_2^2-\hat{\rho}^2)^2}\begin{bmatrix} \hat{\sigma}_2^4N/2 & -\hat{\rho}^2N/2 \\ -\hat{\rho}^2N/2 & \hat{\sigma}_1^4N/2\end{bmatrix}$, $D =\frac{ (\hat{\rho}^2\hat{\rho}'^2+\hat{\sigma}_1^2\hat{\sigma}_2^2\hat{\rho}'^2)\sum_i \tilde{\bm{x}}_i \tilde{\bm{x}}_i^T}{(\hat{\sigma}_1^2\hat{\sigma}_2^2-\hat{\rho}^2)^2} $

\noindent Using Schur's formula, $\begin{bmatrix} A & B \\ C & D \end{bmatrix}^{-1} = \begin{bmatrix} E&F\\ G & H\end{bmatrix}$ where
$$E = (A - BD^{-1}C)^{-1}$$
$$F = -(A-BD^{-1}C)^{-1} BD^{-1}$$
$$G = -D^{-1}C(A-BD^{-1}C)^{-1}$$
$$H = (D-CA^{-1}B)^{-1}$$

\noindent The score statistic is $\hat{d}^T\mathcal{I}\hat{d}$ where
$\hat{d} = (\hat{d}_{\sigma_1^2}, \hat{d}_{\sigma_2^2}, \hat{d}_{\tilde{\bm{\alpha}}}) = (0, 0, \hat{d}_{\tilde{\bm{\alpha}}})$. Therefore, $E$, $F$, $G$ are irrelevant to our score statistic; we only care about the block matrix $H$. Below is step-by-step derivation.

$$A^{-1} = \begin{bmatrix} \frac{\hat{\sigma}_2^4N}{2(\hat{\sigma}_1^2\hat{\sigma}_2^2-\hat{\rho}^2)^2} & -\frac{\hat{\rho}^2N}{2(\hat{\sigma}_1^2\hat{\sigma}_2^2-\hat{\rho}^2)^2} \\
-\frac{\hat{\rho}^2N}{2(\hat{\sigma}_1^2\hat{\sigma}_2^2-\hat{\rho}^2)^2} & \frac{\hat{\sigma}_1^4N}{2(\hat{\sigma}_1^2\hat{\sigma}_2^2-\hat{\rho}^2)^2}\end{bmatrix}^{-1}$$
$$= \frac{4(\hat{\sigma}_1^2\hat{\sigma}_2^2-\hat{\rho}^2)^4}{N^2(\hat{\sigma}_1^4\hat{\sigma}_2^4 - \hat{\rho}^4)}
\begin{bmatrix}
\frac{\hat{\sigma}_1^4N}{2(\hat{\sigma}_1^2\hat{\sigma}_2^2-\hat{\rho}^2)^2} & \frac{\hat{\rho}^2N}{2(\hat{\sigma}_1^2\hat{\sigma}_2^2-\hat{\rho}^2)^2}\\
\frac{\hat{\rho}^2N}{2(\hat{\sigma}_1^2\hat{\sigma}_2^2-\hat{\rho}^2)^2} & \frac{\hat{\sigma}_2^4N}{2(\hat{\sigma}_1^2\hat{\sigma}_2^2-\hat{\rho}^2)^2}
\end{bmatrix} = \frac{2(\hat{\sigma}_1^2\hat{\sigma}_2^2-\hat{\rho}^2)}{N(\hat{\sigma}_1^2\hat{\sigma}_2^2+\hat{\rho}^2)}
\begin{bmatrix}
{\hat{\sigma}_1^4} & {\hat{\rho}^2}\\
{\hat{\rho}^2} & {\hat{\sigma}_2^4}
\end{bmatrix}$$
$$CA^{-1}{B} = \frac{1}{(\hat{\sigma}_1^2\hat{\sigma}_2^2-\hat{\rho}^2)^4} \begin{bmatrix} -\hat{\sigma}_2^2\hat{\rho}\hat{\rho}' & -\hat{\sigma}_1^2\hat{\rho}\hat{\rho}'\end{bmatrix} A^{-1} \begin{bmatrix} -\hat{\sigma}_2^2\hat{\rho}\hat{\rho}' \\ -\hat{\sigma}_1^2\hat{\rho}\hat{\rho}' \end{bmatrix} (\sum_i \tilde{\bm{x}}_i)(\sum_i \tilde{\bm{x}}_i^T)$$
$$ = \frac{1}{(\hat{\sigma}_1^2\hat{\sigma}_2^2-\hat{\rho}^2)^4}\frac{2(\hat{\sigma}_1^2\hat{\sigma}_2^2-\hat{\rho}^2)}{N(\hat{\sigma}_1^2\hat{\sigma}_2^2+\hat{\rho}^2)} \begin{bmatrix} -\hat{\sigma}_2^2\hat{\rho}\hat{\rho}' & -\hat{\sigma}_1^2\hat{\rho}\hat{\rho}' \end{bmatrix} 
\begin{bmatrix}
\hat{\sigma}_1^4 & \hat{\rho}^2 \\ \hat{\rho}^2 & \hat{\sigma}_2^4
\end{bmatrix} 
\begin{bmatrix}
-\hat{\sigma}_2^2\hat{\rho}\hat{\rho}' \\ -\hat{\sigma}_1^2\hat{\rho}\hat{\rho}'
\end{bmatrix}(\sum_i \tilde{\bm{x}}_i)(\sum_i \tilde{\bm{x}}_i^T)$$
$$= \frac{2}{N(\hat{\sigma}_1^2\hat{\sigma}_2^2-\hat{\rho}^2)^3(\hat{\sigma}_1^2\hat{\sigma}_2^2+\hat{\rho}^2)} 
\begin{bmatrix}
\hat{\sigma}_1^4\hat{\sigma}_2^2\hat{\rho}\hat{\rho}' + \hat{\sigma}_1^2\hat{\rho}^3\hat{\rho}' & \hat{\sigma}_2^2\hat{\rho}^3\hat{\rho}' +\hat{\sigma}_1^2\hat{\sigma}_2^4\hat{\rho}\hat{\rho}'
\end{bmatrix}
\begin{bmatrix} -\hat{\sigma}_2^2\hat{\rho}\hat{\rho}' \\ -\hat{\sigma}_1^2\hat{\rho}\hat{\rho}' \end{bmatrix}(\sum_i \tilde{\bm{x}}_i)(\sum_i \tilde{\bm{x}}_i^T)
$$

$$= \frac{2 (\sum_i \tilde{\bm{x}}_i)(\sum_i \tilde{\bm{x}}_i^T)}{N(\hat{\sigma}_1^2\hat{\sigma}_2^2-\hat{\rho}^2)^3(\hat{\sigma}_1^2\hat{\sigma}_2^2+\hat{\rho}^2)}
(\hat{\sigma}_1^4\hat{\sigma}_2^4\hat{\rho}^2\hat{\rho}'^2 + \hat{\sigma}_1^2\hat{\sigma}_2^2\hat{\rho}^4\hat{\rho}'^2 + \hat{\sigma}_1^2\hat{\sigma}_2^2\hat{\rho}^4\hat{\rho}'^2+\hat{\sigma}_1^4\hat{\sigma}_2^4\hat{\rho}^2\hat{\rho}'^2)
$$

$$= \frac{2(\sum_i \tilde{\bm{x}}_i)(\sum_i \tilde{\bm{x}}_i^T)}{N(\hat{\sigma}_1^2\hat{\sigma}_2^2-\hat{\rho}^2)^3(\hat{\sigma}_1^2\hat{\sigma}_2^2+\hat{\rho}^2)}
2\hat{\sigma}_1^2\hat{\sigma}_2^2\hat{\rho}^2\hat{\rho}'^2(\hat{\sigma}_1^2\hat{\sigma}_2^2+\hat{\rho}^2)$$

$$= \frac{4\hat{\sigma}_1^2\hat{\sigma}_2^2\hat{\rho}^2\hat{\rho}'^2(\sum_i \tilde{\bm{x}}_i)(\sum_i \tilde{\bm{x}}_i^T)}{N(\hat{\sigma}_1^2\hat{\sigma}_2^2-\hat{\rho}^2)^3}$$

Then, $H = (D-CA^{-1}B)^{-1}$ = 
$$\left(\frac{\hat{\rho}^2\hat{\rho}'^2+\hat{\sigma}_1^2\hat{\sigma}_2^2\hat{\rho}'^2}{(\hat{\sigma}_1^2\hat{\sigma}_2^2-\hat{\rho}^2)^2}\sum_i \tilde{\bm{x}}_i\tilde{\bm{x}}_i^T -
\frac{4\hat{\sigma}_1^2\hat{\sigma}_2^2\hat{\rho}^2\hat{\rho}'^2(\sum x_i)(\sum x_i^T)}{N(\hat{\sigma}_1^2\hat{\sigma}_2^2-\hat{\rho}^2)^3}\right)^{-1}
$$

$$ = {(\hat{\sigma}_1^2\hat{\sigma}_2^2-\hat{\rho}^2)^3}\left( (\hat{\rho}^2\hat{\rho}'^2 + \hat{\sigma}_1^2\hat{\sigma}_2^2\hat{\rho}'^2)(\hat{\sigma}_1^2\hat{\sigma}_2^2-\hat{\rho}^2)\sum_i \tilde{\bm{x}}_i \tilde{\bm{x}}_i^T -
\frac{4\hat{\sigma}_1^2\hat{\sigma}_2^2\hat{\rho}^2\hat{\rho}'^2}{N}(\sum \tilde{\bm{x}}_i)(\sum \tilde{\bm{x}}_i^T)\right)^{-1}$$

$$ =\frac{(\hat{\sigma}_1^2\hat{\sigma}_2^2-\hat{\rho}^2)^3}{\hat{\rho}'^2}\left( (\hat{\sigma}_1^2\hat{\sigma}_2^2+\hat{\rho}^2)(\hat{\sigma}_1^2\hat{\sigma}_2^2-\hat{\rho}^2)\sum_i \tilde{\bm{x}}_i \tilde{\bm{x}}_i^T -
\frac{4\hat{\sigma}_1^2\hat{\sigma}_2^2\hat{\rho}^2}{N}(\sum \tilde{\bm{x}}_i)(\sum \tilde{\bm{x}}_i^T)\right)^{-1}$$

Consider centering of the covariates so that $\sum_{i=1}^{N} \bm{x}_i = \bm{0}$, and replacing the intercept constant from 1 to $w$ to create $\check{X}$. Then, the above Fisher information matrix becomes block diagonal. 
$$\sum_i \check{\bm{x}}_i \check{\bm{x}}_i^T = \begin{bmatrix}
Nw & \bm{0}\\
\bm{0} & \check{X}^T\check{X}
\end{bmatrix}$$
$$\left(\sum \check{\bm{x}}_i\right)\left(\sum \check{\bm{x}}_i^T\right)= \begin{bmatrix} Nw & \bm{0} \\ \bm{0} & \bm{0}\end{bmatrix}$$
where $w$ is  
\begin{equation}
    w = \frac{(\hat{\sigma}_1^2\hat{\sigma}_2^2+\hat{\rho}^2)(\hat{\sigma}_1^2 \hat{\sigma}_2^2 - \hat{\rho}^2) - \frac{4\hat{\sigma}_1^2\hat{\sigma}_2^2\hat{\rho}^2}{N}}{(\hat{\sigma}_1^2\hat{\sigma}_2^2+\hat{\rho}^2)(\hat{\sigma}_1^2 \hat{\sigma}_2^2 - \hat{\rho}^2)}
    \label{eq:intercept}
\end{equation}
The Fisher Information matrix is equivalent to
\begin{align*}
\hat{I}_{\tilde{\bm{\alpha}}\tilde{\bm{\alpha}}} &= \frac{(\hat{\sigma}_1^2\hat{\sigma}_2^2 - \hat{\rho}^2)^3}{\hat{\rho}'^2}\left((\hat{\sigma}_1^2\hat{\sigma}_2^2+\hat{\rho}^2)(\hat{\sigma}_1^2\hat{\sigma}_2^2-\hat{\rho}^2)\sum_i \check{\bm{x}}_i \check{\bm{x}}_i\right)^{-1}\\
&= \frac{(\hat{\sigma}_1^2\hat{\sigma}_2^2-\hat{\rho}^2)^2}{(\hat{\sigma}_1^2\hat{\sigma}_2^2+\hat{\rho}^2)\hat{\rho}'^2}(\check{X}^T\check{X})^{-1}
\end{align*}

With the modified covariates $\check{X}$ the score statistic can be re-written as 
\begin{equation}
    q_{12} = (\sum_i \check{\bm{x}}_i f_{12,i})(\sum_i \check{\bm{x}}_i \check{\bm{x}}_i^T)^{-1} (\sum_i \check{\bm{x}}_i f_{12,i}) = \phi^T \Psi^{-1} \phi
\end{equation}
where
$$f_{12,i} =  \frac{(\hat{\sigma}_1^2\hat{\sigma}_2^2\hat{\rho}_{12}-\hat{\rho}_{12}^3) + (\hat{\sigma}_1^2\hat{\sigma}_2^2+\hat{\rho}_{12}^2)\hat{u}_{i1}\hat{u}_{i2} - \hat{\sigma}_1^2\hat{\rho}_{12}\hat{u}_{i2}^2-\hat{\sigma}_2^2\hat{\rho}_{12}\hat{u}_{i1}^2}{\sqrt{(\hat{\sigma}_1^2\hat{\sigma}_2^2+\hat{\rho}_{12}^2)(\hat{\sigma}_1^2\hat{\sigma}_2^2-\hat{\rho}_{12}^2)^2}}$$
\noindent Alternatively, collect all $N$ observations by defining $\tilde{X} = \begin{bmatrix} \check{\bm{x}}_1 & \cdots & \check{\bm{x}}_N \end{bmatrix}^T \in \mathbb{R}^{N \times P}$, $f_{12} = \begin{bmatrix} f_{12,1} & \cdots & f_{12,N} \end{bmatrix}^T$, and write
\begin{equation}
q_{12} = f_{12}^T \check{X} (\check{X}^T\check{X})^{-1} \check{X}^T f_{12}
\end{equation}


\vspace{4mm} Following proof from \cite{breusch1979simple}, construct length $N$ vector $g_{12}$ where $g_{12,i} = f_{12,i} + 1$. Then, $q_{12}$ is equal to the explained sum of squares from OLS regression that regresses $X$ from $g_{12}$. Note
$$\sum_{i=1}^{N} f_{12,i} = \frac{N(\hat{\sigma}_1^2\hat{\sigma}_2^2\hat{\rho}_{12} - \hat{\rho}_{12}^3) + (\hat{\sigma}_1^2\hat{\sigma}_2^2+\hat{\rho}_{12}^2)N\hat{\rho}_{12} - \hat{\sigma}_1^2\hat{\rho}_{12}N\hat{\sigma}_2^2 - \hat{\sigma}_2^2\hat{\rho}_{12}N\hat{\sigma}_1^2}{\sqrt{(\hat{\sigma}_1^2\hat{\sigma}_2^2+\hat{\rho}_{12}^2)(\hat{\sigma}_1^2\hat{\sigma}_2^2-\hat{\rho}_{12}^2)^2}} = 0$$
And therefore, $\sum_{i=1}^{N} g_{12,i} = \sum_{i=1}^{N} (f_{12,i} + 1) = N$ and $\bar{g}_{12} = \bm{1}$. 
\begin{align*}
ESS &= 
\sum_{i=1}^{N} (\hat{g}_{12,i} - \bar{g}_{12,i})^2 \\
&= \left(\hat{g}_{12} - \bm{1}\right)^T \left(\hat{g}_{12} - \bm{1}\right) \\
&= \left( \check{X}(\check{X}^T\check{X})^{-1}\check{X}^Tg_{12} - \bm{1}\right)^2 \left( \check{X}(\check{X}^T\check{X})^{-1}\check{X}^Tg_{12} - \bm{1}\right)\\
&= g_{12}^T\check{X}(\check{X}^T\check{X})^{-1}\check{X}^Tg_{12} - 2 \cdot \bm{1}^T\check{X}(\check{X}^T\check{X})^{-1}\check{X}^Tg_{12} + \bm{1}^T \check{X}(\check{X}^T\check{X})^{-1}\check{X}^T\bm{1}\\
&= (f_{12}+\bm{1})^T\check{X}(\check{X}^T\check{X})^{-1}\check{X}^T(f_{12}+\bm{1}) - 2\bm{1}^T\check{X}(\check{X}^T\check{X})^{-1}\check{X}^T(f_{12}+\bm{1}) + \bm{1}^T\check{X}(\check{X}^T\check{X})^{-1}\check{X}^T\bm{1}\\
&= f_{12}^T\check{X}(\check{X}^T\check{X})^{-1}\check{X}^Tf_{12}\\
&= q_{12}
\end{align*}
Consider the regression $g$ onto $X$ and OLS coefficient estimator $\hat{\delta}$. Then, according to \cite{amemiya1977note},
$$\frac{1}{\sqrt{N}} (\hat{\delta} - \delta) \xrightarrow{d} N\left(0, \lim_{N\rightarrow \infty} \frac{1}{N} (\check{X}^T\check{X})^{-1}\tau \right)$$
where
\begin{align*}
\tau &= Var\left(\frac{(\hat{\sigma}_1^2\hat{\sigma}_2^2\hat{\rho}_{12}-\hat{\rho}_{12}^3) + (\hat{\sigma}_1^2\hat{\sigma}_2^2+\hat{\rho}_{12}^2){u}_{i1}{u}_{i2} - \hat{\sigma}_1^2\hat{\rho}_{12}{u}_{i2}^2-\hat{\sigma}_2^2\hat{\rho}_{12}{u}_{i1}^2}{\sqrt{(\hat{\sigma}_1^2\hat{\sigma}_2^2+\hat{\rho}_{12}^2)(\hat{\sigma}_1^2\hat{\sigma}_2^2-\hat{\rho}_{12}^2)^2}}\right)\\
&= E\left(\left(\frac{(\hat{\sigma}_1^2\hat{\sigma}_2^2\hat{\rho}_{12}-\hat{\rho}_{12}^3) + (\hat{\sigma}_1^2\hat{\sigma}_2^2+\hat{\rho}_{12}^2){u}_{i1}{u}_{i2} - \hat{\sigma}_1^2\hat{\rho}_{12}{u}_{i2}^2-\hat{\sigma}_2^2\hat{\rho}_{12}{u}_{i1}^2}{\sqrt{(\hat{\sigma}_1^2\hat{\sigma}_2^2+\hat{\rho}_{12}^2)(\hat{\sigma}_1^2\hat{\sigma}_2^2-\hat{\rho}_{12}^2)^2}}\right)^2\right)
\end{align*}
Using higher moments of multivariate normal: $E((u_{i1}^2u_{i2})^2) = \sigma_1^2\sigma_2^2+\rho_{12}^2$, $E(u_{i1}^4) = 3\sigma_1^4$, $E(u_{i2}^4) = 3\sigma_2^4$, $E(u_{i1}^3u_{i2}) = \sigma_1^2\rho_{12}$, $E(u_{i1}u_{i2}^3) = \sigma_2^2\rho_{12}$,
\begin{align*}
\lim_{N \rightarrow \infty} \tau& = E\left(\left(\frac{(\sigma_1^2\sigma_2^2\rho_{12} - \rho_{12}^3) + (\sigma_1^2\sigma_2^2+\rho_{12}^2)u_{i1}u_{i2} - \hat{\sigma}_1^2\hat{\rho}_{12}{u}_{i2}^2 -  \sigma_1^2\rho_{12}u_{i2}^2}{(\sigma_1^2\sigma_2^2+\rho_{12}^2)(\sigma_1^2\sigma_2^2-\rho_{12}^2)^2}\right)^2\right)\\
&= 1
\end{align*}
Here, we show that if $\hat{\delta}$ follows the normal distribution, then the ESS of the regression $g_{12,i}$ onto $\check{X}$ follows $\chi_P^2$. The ESS of the regression can be written as $Y^TQLQY$ where $L = I - \bm{1}\bm{1}^T/N$ and $Q = \check{X}(\check{X}^T\check{X})^{-1}\check{X}^T$. 
\begin{align*}
    (\hat{g}_{12} - \bar{g}_{12})^T(\hat{g}_{12} - \bar{g}_{12}) &= \left(\hat{g}_{12} - \frac{1}{N} \bm{1}\bm{1}^Tg_{12}\right)^T\left(\hat{g}_{12} - \frac{1}{N} \bm{1}\bm{1}^Tg_{12}\right)\\
    &= \left(\hat{g}_{12} - \frac{1}{N} \bm{1}\bm{1}^T\hat{g}_{12}\right)^T\left(\hat{g}_{12} - \frac{1}{N} \bm{1}\bm{1}^T\hat{g}_{12}\right)\\
    &= \left(Qg_{12} - \frac{1}{N} \bm{1}\bm{1}^TQg_{12}\right)^T\left(Qg_{12} - \frac{1}{N} \bm{1}\bm{1}^TQg_{12}\right)\\
    &= (Qg_{12})^TL^TL(Qg_{12})\\
    &= g_{12}^TQLQg_{12}
\end{align*}
We use the following three intermediary results.
\begin{itemize}
    \item $(Q-L)$ is idempotent because $(I-Q) + (Q-L) + L = I$, and $(I-Q)$ and $L$ are both idempotent. 
    \item $(Q-L)^2 = Q^2 - 2QL + L^2 = Q - 2QL + L$ which is equal to $Q-L$ due to above. Therefore, QL = L, which leads to $QLQLQ = QLLQ = QLQ$. 
    \item rank$(LQ) = tr(LQ) = tr(Q - \bm{1}\bm{1}^TQ/N) = (P+1) - 1 = P$
\end{itemize}
Therefore, we can show that $q_{12}$ follows $\chi_P^2$ distribution asymptotically.
\begin{align*}
q_{12} &= g_{12}^TQLQg_{12} = (X\hat{\delta})^TL(X\hat{\delta}) \text{ where } X\hat{\delta} \rightarrow N(X\delta, Q)
\end{align*}
and if $\delta = 0$, $q_{12}$ follows the chi-squared distribution with degrees of freedom of rank(LQ) = P. 

\section*{Appendix B. Small Sample Correction}
Although the introduced test statistic $q$ asymptotically follows $\chi_1^2$, the approximation has an error in the order of $N^{-1}$ (\cite{harris1985asymptotic}) with finite sample size $N$, and many Monte Carlo experiments show that the test based on it rejects the null hypothesis less frequently than indicated by its nominal size (\cite{godfrey1978testing, griffiths1986monte, honda1988size}). In response, Harris (1985) used Edgeworth expansion to obtain the distribution and moment generating function to order $N^{-1}$ of the test statistic (\cite{harris1985asymptotic}). Building on this expansion, Honda (1986) and Cribari-Neto and Ferrari (2001) proposed corrections to the critical value or to the test statistic that allows better inference even when the sample size is small while preserving the asymptotic properties.\\

Honda (1988) provided a closed-form formula to adjust the critical value in the order of $O(N^{-1})$ to correct the type I error of the test. This adjustment, only depending on the covariate, sample size, and the degrees of freedom, but not on the data, is a cubic function with respect to $C_{\gamma}$. $C_{\gamma}$ is the critical value at the level of type I error $\gamma$, i.e. $P(\chi_{P}^2 \geq C_{\gamma}) = \gamma$. The size-corrected critical value is a cubic function $f$ of the original critical value $C_{\gamma}$ as follows.
\begin{equation}
    \begin{multlined}
f(C_{\gamma})= C_{\gamma} + C_{\gamma}\bigg(\frac{A_3 - A_2 + A_1}{12NP}\bigg) + C_{\gamma}^2\bigg(\frac{A_2 - 2A_3}{(12NP(P+2)}\bigg) +\\
C_{\gamma}^3 \bigg(\frac{A_3}{12NP(P+2)(P+4)}\bigg)
\end{multlined}
\label{eq:hondacorrection}
\end{equation}
\noindent where the scalars $A_1$, $A_2$, and $A_3$ follow the notation of Honda (1988) directly. \cite{breusch1979simple}. 

\vspace{5mm}  
One of the desirable properties of $f$ would be monotonicity, because regardless of sample size, same ordering of the strength of evidence against the null is expected and is interpretable. Below, we show that monotonicity asymptotically holds and is almost always true in practice. The derivative of $f(C_{\gamma})$ is 
$$f'(C_{\gamma}) = \frac{A_3}{12NP}\left( \frac{A_3-A_2+A_1+12NP}{A_3} + 
 \frac{2(A_2-2A_3)}{(P+2)A_3}C_{\gamma} + \frac{3}{(P+2)(P+4)}C_{\gamma}^2 \right)$$
$A_3$ is strictly positive by definition, and we can solve the above quadratic equation to see in which case the derivative is positive (\cite{cribari1995improved}). In other words, we study whether the following discriminant is complex.
$$\sqrt{\left(\frac{2(A_2-2A_3)^2}{(P+2)A_3}\right)^2 - 4\cdot\frac{3A_3(A_3-A_2+A_1)}{(P+2)(P+4)A_3} - 4\cdot
\frac{3 \cdot 12NP}{(P+2)(P+4)}}$$
The first two terms inside the square root are $O(1)$ and the last term is $O(N)$, so we can see that the discriminant becomes complex quickly as $N$ increases.\\


\noindent We aim to adjust the test statistic so that the overall shape of the null distribution is closer to $\chi_{P}^2$. We assume a large enough sample size for monotonicity of $g$ and define the inverse function of $g$ to propose the new adjusted test statistic $\tilde{q}$ = $f^{-1}(q)$
$$\gamma = P(\chi_{P}^2 \geq C_{\gamma}) = P(q \geq f(C_{\gamma})) = P(f^{-1}(q) \geq C_{\gamma}).$$
The test statistic $q$ is replaced by solving the following equation
$$q: q - f(C_{\gamma}) = 0$$
which is guaranteed to be unique by the monotonicity of $f$. The cubic equation can be solved given the covariate $X$.

\section*{Appendix C. Derivation of the distribution of $d_1$}

Here, we derive the distribution of $d_1 = q_{12} + q_{13} + \cdots q_{1K}$ under the global null hypothesis (\ref{eq:globalnull}). \\

We first scale and orthogonalize $\check{X}$ so that $\sum \check{x}_{ip}^2 = 1$ and $\sum_i \check{x}_{ip_1}\check{x}_{ip_2} = 0$ for all $p_1 \neq p_2$ in $1, \cdots, P$. This transformation does not affect the error $u$ or the residuals $\hat{u}$ and allows to replace $(\check{X}^T\check{X})^{-1}$ with $\frac{1}{N} I$. Then, the score statistic can be re-written as below. 
\begin{align*}
q_{12} &= \frac{1}{N}\left(\sum_{i=1}^{N} \check{\bm{x}}_i^Tf_{12,i}\right)^T \left(\sum_{i=1}^{N} \check{\bm{x}}_i^Tf_{12,i}\right) \\
&= \sum_{p=1}^{P} \left(\sum_i \check{x}_{ip} f_{12,i}\right)^2\\
&= \sum_{p=1}^{P} \left( \sum_i \frac{1}{\sqrt{N}} \frac{1}{W_{12}}  V_{12,ip}\right)^2
\end{align*}
where
\begin{align*}
&W_{12} = \sqrt{(\hat{\sigma}_1^2\hat{\sigma}_2^2+\hat{\rho}_{12}^2)(\hat{\sigma}_1^2\hat{\sigma}_2^2-\hat{\rho}_{12}^2)^2}\\
&V_{12,ip} = \check{x}_{ip}\left((\hat{\sigma}_1^2\hat{\sigma}_2^2\hat{\rho}_{12}-\hat{\rho}_{12}^3) + (\hat{\sigma}_1^2\hat{\sigma}_2^2+\hat{\rho}_{12}^2)\hat{u}_{i1}\hat{u}_{i2} - \hat{\sigma}_1^2\hat{\rho}_{12}\hat{u}_{i2}^2-\hat{\sigma}_2^2\hat{\rho}_{12}\hat{u}_{i1}^2\right).
\end{align*}

Then,
\begin{align*}
    d_1 &= \sum_{k=2}^{K} q_{1k}\\
    &= \sum_{k=2}^{K} \sum_{p=1}^{P} \left( \sum_{i=1}^{N} \frac{1}{\sqrt{N}} \frac{1}{W_{1k}}  V_{1k,ip}\right)^2\\
    &= \sum_{p=1}^{P} \sum_{k=2}^{K} r_{1k,p}^2 = \sum_{p=1}^{P} \|\bm{r}_{1,p}\|^2
\end{align*}
where $\bm{r}_{1,p} = \begin{bmatrix} r_{12,p} & r_{13,p} & \cdots & r_{1K,p}\end{bmatrix}^T$. Below, we show that $\bm{r}_{1,p}$ asymptotically follows multivariate normal distribution $\mathcal{N}(\bm{0}, H_1)$, where the covariance matrix $H_1$ has 1 in the diagonals and $\eta_{k\ell}$ for $k \neq \ell$ in non-diagonals. We derive $\eta_{k\ell}$ below in a closed form as well.\\

First, $r_{1k,p} = \sum_{i=1}^{N} \frac{1}{\sqrt{N}} \frac{1}{W_{1k}} V_{1k,ip}$ asymptotically follows standard normal distribution. $W_{1k}$ converges in probability to a constant. Using the definitions, $\hat{\sigma}_1^2 = \frac{1}{N} \sum_{i=1}^{N} \hat{u}_{i1}^2 \xrightarrow{P} \sigma_1^2$, $\hat{\sigma}_2^2 = \frac{1}{N} \sum_{i=1}^{N} \hat{u}_{i2}^2 \xrightarrow{P} \sigma_2^2$, and
$\hat{\rho}_{12} = \frac{1}{N} \sum_{i=1}^{N} \hat{u}_{i1}\hat{u}_{i2} \xrightarrow{P} \rho_{12}$. 
Continuous Mapping Theorem shows that
$$\frac{1}{W_{12}} \xrightarrow{P} \frac{1}{\sqrt{({\sigma}_1^2{\sigma}_2^2+{\rho}_{12}^2)({\sigma}_1^2{\sigma}_2^2-{\rho}_{12}^2)}}$$
$V_{1k,ip}$ is asymptotically independent across $i = 1, \cdots, N$ so that we can use Central Limit Theorem in $\frac{1}{\sqrt{N}} \sum_{i=1}^{N} V_{1k,ip}$. $V_{1k,ip}$ is a function of the regression residuals $\hat{u}_{i1}$ and $\hat{u}_{i2}$ that are not independent across $i$ in finite sample $N$. The residual can be written as
$$\hat{u}_1 = (I-\check{X}(\check{X}^T\check{X})^{-1}\check{X}^T)u_1 = (I-\frac{1}{N} \check{X}\check{X}^T)u_1$$
where $u_1$ is i.i.d normal error vector. The non-diagonal term of the matrix $\left(I - \frac{1}{N} \check{X}\check{X}^T\right)$ is $ -\frac{1}{N} \check{\bm{x}}_i^T\check{\bm{x}}_j$. Meanwhile, asymptotically, the non-diagonal elements converge in probability to 0 under mild conditions about the covariates and their dimension $P$, i.e. when P is finite and fixed, and every element of matrix $X$ is finite and fixed. Therefore, asymptotically, $\hat{u}_1 = u_1$ and $\hat{u}_2 = u_2$, making the residuals asymptotically independent across $i$. 
\begin{align*}
E(\lim_{N \rightarrow \infty}\sum_{i=1}^{N} V_{1k,ip}) 
&=  E( \lim_{N \rightarrow \infty}\sum_{i=1}^{N} \check{x}_{ip}\left((\hat{\sigma}_1^2\hat{\sigma}_2^2\hat{\rho}_{1k}-\hat{\rho}_{1k}^3) +  (\hat{\sigma}_1^2\hat{\sigma}_2^2+\hat{\rho}_{1k}^2)\hat{u}_{i1}\hat{u}_{i2} - \hat{\sigma}_1^2\hat{\rho}_{1k}\hat{u}_{i2}^2-\hat{\sigma}_2^2\hat{\rho}_{1k}\hat{u}_{i1}^2\right))\\
&= \lim_{N \rightarrow \infty}\sum_{i=1}^{N} \check{x}_{ip} E(\rho_{1k}(\sigma_1^2\sigma_k^2 - \rho_{1k}^2) + (\sigma_1^2\sigma_k^2+\rho_{1k}^2)\hat{u}_{i1}\hat{u}_{i2} - \sigma_1^2\rho_{1k}\hat{u}_{i2}^2 - \sigma_k^2\rho_{1k}\hat{u}_{i1}^2)\\
&=\lim_{N \rightarrow \infty}\sum_{i=1}^{N} \check{x}_{ip}(\rho_{1k}(\sigma_1^2\sigma_k^2 - \rho_{1k}^2) + (\sigma_1^2\sigma_k^2+\rho_{1k}^2)\rho_{1k} - \sigma_1^2\rho_{1k}\sigma_k^2 - \sigma_k^2\rho_{1k}\sigma_1^2)\\
&= 0
\end{align*}
\begin{align*}
&Var(\lim_{N \rightarrow \infty} \sum_{i=1}^{N} V_{1k,ip})\\
&= \sum_{i=1}^{N}  E(\lim_{N \rightarrow \infty} \check{x}_{ip}^2\left((\hat{\sigma}_1^2\hat{\sigma}_2^2\hat{\rho}_{1k}-\hat{\rho}_{1k}^3) +  (\hat{\sigma}_1^2\hat{\sigma}_2^2+\hat{\rho}_{1k}^2)\hat{u}_{i1}\hat{u}_{i2} - \hat{\sigma}_1^2\hat{\rho}_{1k}\hat{u}_{i2}^2-\hat{\sigma}_2^2\hat{\rho}_{1k} \hat{u}_{i1}^2\right)^2)\\
&=N(\rho_{1k}^2(\sigma_1^2\sigma_k^2-\rho_{1k}^2)^2 + (\sigma_1^2\sigma_k^2+\rho_{1k}^2)^2(\sigma_1^2\sigma_k^2+2\rho_{1k}^2) + \sigma_1^4\rho_{1k}^2\cdot 3\sigma_k^4 + \sigma_k^4\rho_{1k}^2\cdot 3\sigma_1^4 \\
&\hspace{5mm}+2\rho_{1k}(\sigma_1^2\sigma_k^2-\rho_{1k}^2)(\sigma_1^2\sigma_k^2+\rho_{1k}^2)\cdot \rho_{1k} - 2\rho_{1k}(\sigma_1^2\sigma_k^2-\rho_{1k}^2) - \sigma_1^2\rho_{1k}\sigma_k^2-2\rho_{1k}(\sigma_1^2\sigma_k^2-\rho_{1k}^2)\sigma_k^2\rho_{1k}\cdot \sigma_1^2 \\
& \hspace{5mm} - 2\sigma_1^2\rho_{1k}(\sigma_1^2\sigma_k^2+\rho_{1k}^2)\cdot  3\sigma_k^2\rho_{1k}  - 2\sigma_k^2\rho_{1k}(\sigma_1^2\sigma_k^2 +\rho_{1k}^2)\cdot 3\sigma_1^2\rho_{1k} + 2\sigma_1^2\sigma_k^2\rho_{1k}^2 \cdot (\sigma_1^2\sigma_k^2+2\rho_{1k}^2) )\\
&= N (\sigma_1^2\sigma_k^2+\rho_{1k}^2)({\sigma}_1^2{\sigma}_2^2-\rho_{1k}^2)^2
\end{align*}

and therefore,
\begin{equation}
\label{wandvstandardnormal}
\frac{1}{W_{1k}}\left( \frac{1}{\sqrt{N}} \sum_{i=1}^{N}V_{1k,ip}\right) \rightarrow \mathcal{N}(0, 1)
\end{equation}

\noindent Next, we derive the asymptotic correlation between $r_{1k,p}$ and $r_{1\ell,p}$ by deriving $cov(\lim_{n \rightarrow \infty} V_{1k,ip}, \lim_{n \rightarrow \infty} V_{1\ell, i})$ and hence derive the non-diagonal elements of $H_1$, $\eta_{k,\ell}$. We first derive the intermediary result of $h_{k,\ell}$:\\
\begin{align*}
&h_{k,\ell} = \lim_{N \rightarrow \infty} cov\left(\sum_{i=1}^{N} V_{1k,ip}, \sum_{i=1}^{N} V_{1\ell, ip}\right)\\
&= \lim_{N \rightarrow \infty} E(\sum_{i=1}^{N} V_{1k,ip} \sum_{i=1}^{N}V_{1\ell,ip})  \\
&= \lim_{N \rightarrow \infty} E(\sum_{i=1}^{N} V_{1k,ip}V_{1\ell,ip})\hspace{5mm}  \textit{(asymptotic independence across $i$)}\\
&= \lim_{N \rightarrow \infty} \sum_{i=1}^{N}\check{x}_{ip}^2E(\left((\hat{\sigma}_1^2\hat{\sigma}_2^2\hat{\rho}_{12}-\hat{\rho}_{12}^3) + (\hat{\sigma}_1^2\hat{\sigma}_2^2+\hat{\rho}_{12}^2)\hat{u}_{i1}\hat{u}_{i2} - \hat{\sigma}_1^2\hat{\rho}_{12}\hat{u}_{i2}^2-\hat{\sigma}_2^2\hat{\rho}_{12}\hat{u}_{i1}^2\right))\cdot \\
& \hspace{13mm} \left((\hat{\sigma}_1^2\hat{\sigma}_2^2\hat{\rho}_{12}-\hat{\rho}_{12}^3) + (\hat{\sigma}_1^2\hat{\sigma}_2^2+\hat{\rho}_{12}^2)\hat{u}_{i1}\hat{u}_{i2} - \hat{\sigma}_1^2\hat{\rho}_{12}\hat{u}_{i2}^2-\hat{\sigma}_2^2\hat{\rho}_{12}\hat{u}_{i1}^2\right)\\
&= N(2\sigma_1^4\sigma_k^2\sigma_{\ell}^2\rho_{1k}\rho_{1\ell} + 3\sigma_1^2\sigma_{\ell}^2\rho_{1k}^3\rho_{1\ell} + 3\sigma_1^2\sigma_k^2\rho_{1k}\rho_{1\ell}^3 - 2\sigma_1^4\sigma_k^4\sigma_{\ell}^2\rho_{1\ell}  \\
& \hspace{6mm} - \sigma_1^4\sigma_k^2\rho_{1\ell}^2\rho_{k\ell} - \sigma_1^2\sigma_k^4\rho_{1\ell}^3 + \sigma_1^6\sigma_k^2\sigma_{\ell}^2\rho_{k\ell} - 2\sigma_1^2\sigma_k^2\sigma_{\ell}^2\rho_{1k}^2\rho_{1\ell} - \sigma_k^2\rho_{1k}^2\rho_{1\ell}^3 \\
& \hspace{6mm}- \sigma_1^4\sigma_{\ell}^2\rho_{1k}^2\rho_{k\ell} + 2\rho_{1k}^3\rho_{1\ell}^3 - \sigma_1^4\sigma_k^2\sigma_{\ell}^2\rho_{1k}^2-\sigma_1^2\sigma_k^2\rho_{1k}^2\rho_{1\ell}^2+2\sigma_1^4\rho_{1k}\rho_{1\ell}\rho_{k\ell}^2)
\end{align*}
This leads to the non-diagonal elements of $H_1$,
\begin{equation}
\eta_{k,\ell} = \frac{h_{k,\ell}}{\sqrt{({\sigma}_1^2{\sigma}_k^2+{\rho}_{1k}^2)({\sigma}_1^2{\sigma}_k^2-{\rho}_{1k}^2)^2({\sigma}_1^2{\sigma}_{\ell}^2+{\rho}_{1\ell}^2)({\sigma}_1^2{\sigma}_{\ell} ^2-{\rho}_{1\ell}^2)^2}}
\label{eq:eta}
\end{equation}
Therefore, we have shown that 
$\bm{r}_{1,p} \xrightarrow{D} \mathcal{N}(\bm{0}, H_1)$ with closed-form definition of $H_1$. Finally, we derive the null distribution of $d_1 = \sum_{p=1}^{P}\|\bm{r}_{1,p}\|_2^2$. Let $H_1 = U_1\Lambda_1U_1^T$ be the eigen-decomposition of $H_1$, where $\Lambda$ is a diagonal matrix with elements $\lambda_{12}, \cdots, \lambda_{1K}$. Due to the orthogonal invariance of L2 norm, $d_1 = \| \bm{r}_{1,p}\|^2 = \|U_1\bm{r}_{1,p}\|^2$. Then, 
$$U_1\bm{r}_{1,p} \sim \mathcal{N}(0, \Lambda_1)$$
Each component of the left hand side is independent with known variance $\lambda_{1k}$ for $k = 2, \cdots, K$. The sum of the square of normal variables can be written as a sum of independent gamma variables:
$$\sum_{p=1}^{P} r_{1k,p}^2 \sim \Gamma\left(P/2, \lambda_{1k}/2\right)$$
$$d_1 = \sum_{k=2}^{K}\sum_{p=1}^P  r_{1k,p}^2 \sim \sum_{k=2}^{K} \Gamma\left( \frac{P}{2}, \frac{\lambda_{1k}}{2}\right)$$ 
This concludes the proof of Proposition \ref{prop:d}.

\section*{Appendix D. Data}
We use the genotype and normalized gene expression data from GTEx V6p release (\cite{lonsdale2013genotype}) to apply the proposed method to the African American samples. The data has been pre-processed by GTEx consortium as explained in the GTEx portal (https://gtexportal.org). In order to select African Americans from the available samples, we first inferred the local ancestry of the samples who identified themselves as European Americans or African Americans and verified that their genetic ancestry is consistent with the reported information. For local ancestry inference, we use the software LAMP that reaches as high as 98\% accuracy level for distinguishing YRI and CEU ancestry (\cite{pacsaniuc2009imputation}). We also used the reference minor allele frequency from pure populations from 1000 Genome Project. For the initial setting of hyperparameters in LAMP, we use 7 for the number of generations of admixture, 0.2 and 0.8 for the initial proportion of CEU and YRI population, and $10^{-8}$ for recombination rate, but the results are robust to these settings. LAMP returns local ancestry at each SNP as the count of African chromosomes (0, 1, or 2) at each locus, and we use the SNP closest to the center of the gene to represent the local ancestry of the entire gene.  Around 92\% of the genes show no recombination event in all of the subjects, and less than 3\% of the genes have more than one individual with ancestry switch within the gene, so we believe this is a valid approximation. We compute the global ancestry by averaging the inferred local ancestry, and this estimate is cross-checked with principal component analysis which can effectively cluster the subjects into sub-populations (\cite{pritchard2000inference}). We also include pure YRI and CEU population for PCA, and most African Americans lie strictly between the YRI and CEU population showing a two-way admixture between pure Europeans and pure Africans. We observed some outliers that were not placed between pure populations, and so we removed them. We also observed some self-identified Europeans whose genetic ancestry is more than 10\% African, and we include them in our analysis as African Americans. 

\end{document}